\documentclass[preprint,prd,superscriptaddress,nofootinbib]{revtex4-2}

\usepackage{natbib}
\usepackage{xcolor}
\usepackage[utf8]{inputenc}
\usepackage{amsfonts,bm}
\usepackage{amsmath,amssymb,graphicx,textcomp}
\usepackage{hyperref}
\usepackage[caption=false]{subfig}

\newcommand{\kl}{K_{\textrm{L}}}
\newcommand{\kll}{\kl\rightarrow\mu^+\mu^-}

\newcommand{\chw}{\mathcal{H}_{\textrm{W}}}
\newcommand{\lla}{\left\langle}
\newcommand{\rra}{\right\rangle}

\DeclareFontFamily{U}{mathx}{}
\DeclareFontShape{U}{mathx}{m}{n}{<-> mathx10}{}
\DeclareSymbolFont{mathx}{U}{mathx}{m}{n}
\DeclareMathAccent{\widecheck}{0}{mathx}{"71}

\begin{document}

\newcommand*{\CU}{Physics Department, Columbia University, New York City, New York 10027, USA}\affiliation{\CU}
 
\title{Calculating the two-photon exchange contribution to $K_L\to\mu^+\mu^-$ decay}
\author{En-Hung Chao}\affiliation{\CU}
\author{Norman Christ}\affiliation{\CU}

\date{June 11, 2024}

\begin{abstract}
We present a theoretical framework within which both the real and imaginary parts of the complex, two-photon exchange amplitude contributing to $\kl\to\mu^+\mu^-$ decay can be calculated using lattice QCD.  The real part of this two-photon amplitude is of approximately the same size as that coming from a second-order weak strangeness-changing neutral-current process.  Thus a test of the standard model prediction for this second-order weak process depends on an accurate result of this two-photon amplitude.  A limiting factor of our proposed method comes from low-energy three-particle $\pi\pi\gamma$ states.  The contribution from these states will be significantly distorted by the finite volume of our calculation -- a distortion for which there is no available correction.  However, a simple estimate of the contribution of these three-particle states suggests their contribution to be at most a few percent allowing their neglect in a lattice calculation with a 10\% target accuracy.
\end{abstract}

\maketitle


\newpage

\section{Introduction}
\begin{figure}[h]
\includegraphics[scale=1]{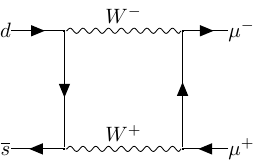}
\hspace{16pt}
\includegraphics[scale=1]{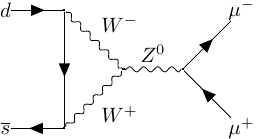}
\hspace{16pt}
\includegraphics[scale=1]{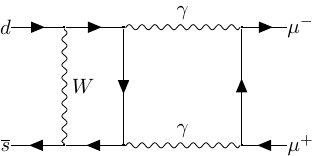}
\caption{The one-loop SD (left and middle) and the LD2$\gamma$ (right) contributions to the $K_{\rm L}\rightarrow\mu^+\mu^-$ decay amplitude.}
\label{fig:pert}
\end{figure}

Despite its rarity, the decay of a long-lived kaon into a pair of charged muons (KL2$\mu$) has been measured to a good precision since the E871 experiment at the Brookhaven National Laboratory two decades ago~\cite{E871:2000wvm}, leading to a branching ratio of $6.84(11)\times 10^{-9}$~\cite{Workman:2022ynf}.
In the standard model, this strangeness-changing weak neutral current process starts at one-loop order and requires exchanging two W bosons or two W and one Z boson in a perturbative electroweak expansion (see Fig.~\ref{fig:pert}).
Such contributions are commonly referred to as the short-distance (SD) part of the decay amplitude.  These are of special interest because of their sensitivity to physics at the electroweak scale and above.
The calculation of the SD part has been carried out in perturbative Quantum Chromodynamics (QCD) using renormalization group techniques~\cite{Buchalla:1993wq}.
An improved estimate of the SD part with the charm-quark contribution evaluated at next-to-next-to-leading order in $\alpha_s$ gives a contribution to the $\kl\to\mu^+\mu^-$ branching ratio of $\textrm{Br}(\kl\rightarrow\mu^+\mu^-)_{\rm SD} = 0.79(12)\times 10^{-9}$~\cite{Gorbahn:2006bm}.

However, a direct comparison of this SD contribution to the experimental result is hindered by a sizeable long-distance (LD) contribution coming at two-loop order in electroweak perturbation from the exchange of two photons and a single W boson (LD2$\gamma$), see Fig.~\ref{fig:pert}.
Although the LD2$\gamma$ contribution starts at two-loop level in perturbation theory, the relative size of the fine-structure constant $\alpha_{\rm QED}$ and the Fermi constant $G_{\rm F}$, makes this two-loop $O(\alpha_{\rm QED}^2 G_{\rm F})$ contribution as important as the one-loop SD contribution. The importance of this two-loop LD contribution can be seen if the optical theorem is used to calculate the absorptive (imaginary) part of the KL2$\mu$ decay amplitude which comes from the on-shell intermediate $\gamma\gamma$ state.  Using the measured $\kl\rightarrow\gamma\gamma$ decay rate, one finds that the absorptive part coming from the 2$\gamma$ channel alone gives a branching ratio of $\textrm{Br}(\kl\rightarrow\mu^+\mu^-)_{\rm{abs}} = 6.59(5)\times 10^{-9}$ providing 98\% of the total decay rate found in experiment~\cite{Martin:1970ai}.  

The sizes of these various contributions to the $K_L\to\mu^+\mu^-$ decay amplitude are compared in Table~\ref{tab:comparison} where the experimental prediction for the magnitude of the real part is obtained from the known imaginary part and the total decay rate.  The complex decay amplitude $\mathcal{A}_{\kl\mu\mu}(s)$ appearing in that table is defined by
\begin{equation}
\langle\mu^+(\textbf{k}_+, s)\mu^+(\textbf{k}_-,s)|\mathcal{H}_{\rm W}(0)|K_L(\textbf{P})\rangle
  = \mathcal{A}_{K_L\mu\mu}(s)\,. 
\label{eq:KLmm-def}
\end{equation}
Here $\textbf{k}_\pm$ are the three-momenta of the final-state muons, $\textbf{P}$ the three-momentum of the kaon in the kaon rest system, and $\mathcal{H}_{\rm W}$ the effective $\Delta S =1$ weak Hamiltonian.
The variable $s=\pm\frac{1}{2}$ is the common helicity of both muons.  All three states whose product appears in Eq.~\eqref{eq:KLmm-def} follow the same covariant normalization conventions:
\begin{equation}
\langle \textbf{k}|\textbf{k}'\rangle = 2 E(\textbf{k}) (2\pi)^3 \delta^3(\textbf{k} -\textbf{k}').
\end{equation}
The amplitude $\mathcal{A}_{K_L\mu\mu}(s)$ is an odd function of $s$.  With these conventions the $K_L\to\mu^+\mu^-$ decay rate is given by:
\begin{equation}
\Gamma_{K_L\to\mu^+\mu^-} = \frac{\beta_\mu}{8\pi M_K}\left|\mathcal{A}_{K_L\mu\mu}(s) \right|^2\,,
\end{equation} 
where $\beta_\mu = \sqrt{1-4m_\mu^2/M_K^2}$ is the velocity of each emitted muon.  
\begin{table}[h]
\centering
\begin{tabular}{l|c|c}
Source & Re$\mathcal{A}_{K_L\mu\mu}(s)$ & Im$\mathcal{A}_{K_L\mu\mu}(s)$  \\
\hline \hline
Experiment      & $\pm0.232(0.021)$   & $1.078(0.005)$ \\
Short distance &  0.375(0.028)  & 0  \\
On-shell $\pi\pi\gamma$ state  & \textemdash  & -0.00146(2) 
\end{tabular}
\caption{Comparison among some of the contributions to the real and imaginary parts of $\mathcal{A}_{K_L\mu\mu}(s)$ as well as with the experimental decay amplitude.  For the experimental case, the imaginary amplitude is taken from the optical theorem while the real part is computed from the decay rate and that imaginary part.  All numbers are in units of $10^{-9}$ MeV.  (Throughout this paper we have not adopted a sign convention that would give meaning to the overall signs of the estimates presented.)}
\label{tab:comparison}
\end{table}
We also include in Table~\ref{tab:comparison} the optical theorem contribution of the $\pi\pi\gamma$ state estimated below in a simple model calculation in which hadronic structure is neglected.  

Because of the similar sizes of the dispersive LD2$\gamma$ contribution and the SD part of interest, a first-principles calculation of this LD2$\gamma$ contribution is required.
Earlier estimates of this dispersive part are based on low-energy QCD phenomenology~\cite{DAmbrosio:1997eof, Knecht:1999gb, Isidori:2003ts}~\footnote{See Ref.~\cite{Cirigliano:2011ny} for a more exhaustive list of references.}.
The crucial element in these analyses is the off-shell $\kl\rightarrow\gamma^*\gamma^*$ transition form factor, which has been recently re-evaluated based on dispersion-relation arguments~\cite{Hoferichter:2023wiy}.

The aim of this paper is to further develop the approach proposed in Refs.~\cite{Christ:2020bzb, Christ:2020dae} and previously applied to the decays $\pi^0\to e^+e^-$~\cite{Christ:2022rho} and $\kl\to\gamma\gamma$~\cite{Zhao:2022pbs} to permit the more challenging calculation of the dispersive part of this LD2$\gamma$ contribution using lattice QCD, the only established systematically-improvable method to study low-energy QCD from first principles.
Analogous to the lattice QCD calculation of the hadronic light-by-light contribution to the anomalous magnetic moment of the muon~\cite{Green:2015sra, Blum:2017cer}, we treat QED in the continuum and infinite-volume (QED${}_\infty$) and expect exponentially-suppressed finite-size effects introduced by the finite-volume treatment of the QCD amplitude.
This formalism does not rely on any parametrization of the off-shell $K_{\rm L}\rightarrow\gamma^*\gamma^*$ form factor, but care must be taken to remove unphysical, exponentially-growing contributions from intermediate states which are less energetic than the kaon.

Such intermediate states less energetic than the initial kaon require special attention in the case of this LD2$\gamma$ amplitude.  While the unphysical contributions of the discrete finite-volume states less energetic than the kaon which lead to exponential growth as the time extent of the lattice calculation increases can be simply removed, it is only in the case of one- or two-particle states that the power-law finite-volume errors associated with these states can be controlled~\cite{Christ:2015pwa}.  For the calculation of the LD2$\gamma$ amplitude proposed here there is a three-particle state composed of a two-pion, finite-volume QCD eigenstate combined with an infinite-volume recoiling photon that may be less energetic than the initial kaon.  While such a state could be identified and its unphysical, exponentially growing contribution subtracted, the finite-volume errors remaining from such a treatment of this three-particle state are not known. In this paper we do not propose a solution to this problem.  Instead we argue that the contribution from such $\pi\pi\gamma$ states is sufficiently small that any needed finite-volume corrections can be safely ignored.

This paper is organized as follows. 
In Sect.~\ref{sec:eucl}, we explain how to extract the complex KL2$\mu$ decay amplitude from a calculation in Euclidean space and how the difficulties introduced by the presence of one- and two-particle intermediate states less energetic than the initial kaon can be addressed.  In Sect.~\ref{sec:pi-pi-gamma}, we estimate the total contribution to the KL2$\mu$ amplitude coming from $\pi\pi\gamma$ states in this troublesome kinematic region and conclude that their total contribution is at or below the few-percent level.  Concluding remarks are made in Sect.~\ref{sec:conclu} and some additional details are given in the two appendices.

\section{Euclidean space evaluation}\label{sec:eucl}
\begin{figure}[h]
\centering
\includegraphics[width=0.5\textwidth]{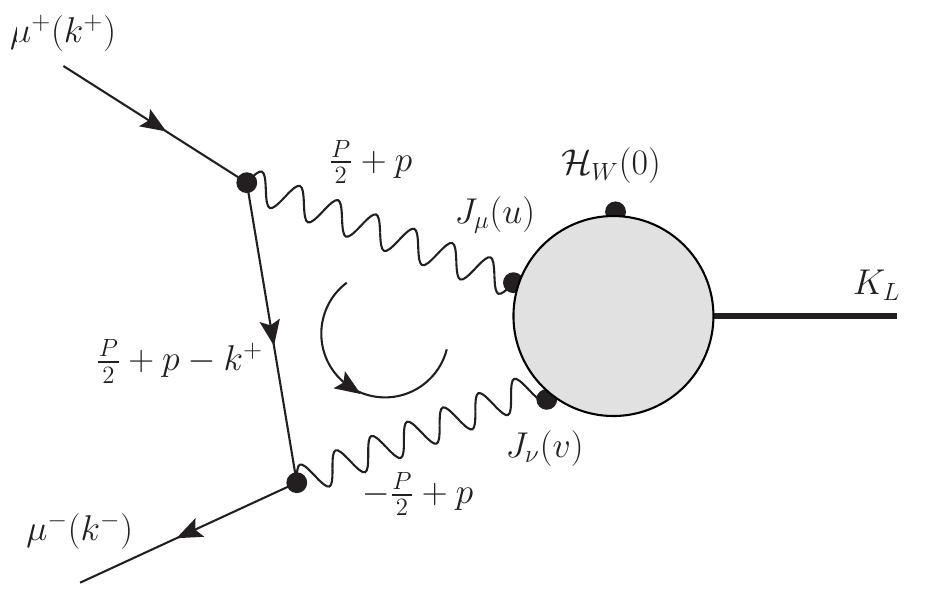}
\caption{Feynman graph representing the amplitude that appears in Eq.~\eqref{eq:K-MS}.  The momentum routing used in that equation is also shown.}
\label{fig:Feynman}
\end{figure}

In this section we develop the method that allows the complex Minkowski-space two-photon exchange contribution to the $K_L\to\mu^+\mu^-$ decay amplitude to be obtained from a Euclidean-space, four-point Green's function which can be computed using lattice QCD.  This is an extension of the method presented in Ref.~\cite{Christ:2022rho} where a similar result was obtained for the two-photon exchange contribution to $\pi^0\to e^+ e^-$ decay.  The current situation is more complex because of the presence in the decay amplitude of a third hadronic operator, the effective weak interaction $\Delta S=1$, four-quark Hamiltonian density $\mathcal{H}_W$.  We begin with a general expression for this two-photon-exchange $K_L\to\mu^+\mu^-$ amplitude evaluated in fifth-order electroweak perturbation theory:
\begin{eqnarray}
\mathcal{A}_{ss'}(k^+,k^-) &=& e^4\int\!\! d^4 u \; \int\!\! d^4 v \;\int\!\! \frac{d^4 p}{(2\pi)^4} \;  
\frac{e^{-i\left(\frac{P}{2}+p\right)u}  e^{-i\left(\frac{P}{2}-p\right)v}}{(\frac{P}{2}+p)^2+m_\gamma^2-i\varepsilon}
\frac{1}{(\frac{P}{2}-p)^2+m_\gamma^2-i\varepsilon}
\label{eq:K-MS} \\
&& \hskip -0.5 in\times\frac{\overline{u}_s(k^-)\gamma_\nu\{\gamma\cdot(\frac{P}{2}+p-k^+) +m_\mu\}\gamma_\mu v_{s'}(k^+)}
      {(\frac{P}{2}+p-k^+)^2 +m_\mu^2-i\varepsilon} \left\langle 0\left|\textrm{T}\left\{ J_\mu(u) J_\nu(v) \mathcal{H}_W(0)\right\}\right|K_L\right\rangle, 
\nonumber
\end{eqnarray}
where we have used the space-time translation symmetry of this on-shell decay amplitude to set to zero the location of $\mathcal{H}_W$.  Here $P$ is the four-momentum of the kaon.  We will work in the kaon's rest system with  $P=(M_K,\textbf{0})$.  We have also introduced a photon mass $m_\gamma$ to make it easier to recognize a photon energy when it appears and assume that the two roles of the symbol $\mu$, identifying the muon mass $m_\mu$ and the space-time component of the four-vector $\gamma_\mu$ can be distinguished.  The Feynman graph shown in Fig.~\ref{fig:Feynman} represents the amplitude and momentum routing shown in Eq.~\eqref{eq:K-MS}.

As in the case of the $\pi^0\to e^+ e^-$ decay we will use Cauchy's theorem to deform the integrals over the time coordinates $u_0$ and $v_0$ from the real axis to the imaginary axis so that the hadronic amplitude appearing as the right-most factor on the right-hand side of Eq.~\eqref{eq:K-MS} is evaluated in Euclidean space, allowing it to be calculated using lattice QCD.  As is well known, such a Wick rotation to Euclidean space is possible for a general $N^{th}$-order decay amplitude of the form
\begin{equation}
\mathcal{A}_{F,D,N} = \prod_{j=1}^{N-1}\left\{\int_{-\infty}^\infty d t_j\right\}
             \left\langle F\left|T\left\{H_{N-1}(t_{N-1})H_{N-2}(t_{N-2})\ldots H_1(t_1) H_0(t_0=0)\right\}\right|D\right\rangle
\end{equation}
for a given time ordering, $t_{\sigma(N-1)} \ge t_{\sigma(N-2)} \ge \ldots t_{\sigma(1)} \ge t_{\sigma(0)}$ if the energy of each possible intermediate state $|n\rangle$ which can be inserted between any pair of adjacent operators as in the expression $H_{\sigma(j)} |n\rangle\langle n|H_{\sigma(j-1)}$ has an energy $E_n$ which is greater than the common energy of the decaying state $|D\rangle$ and the final state $| F \rangle$.  Here $\sigma(j)$ is a permutation of the $N$ integers $0,1,\ldots N-1$ that identifies a particular time ordering.  For completeness, we provide a demonstration of this result in Appendix~\ref{sec:M=E}.

This result cannot be used in the present case because the total energy $E_{\gamma\gamma}$ of the intermediate two-photon state can be smaller than $M_K$.  However, we will show in this section that the Wick rotation of only the $u_0$ and $v_0$ contours is possible provided that the energy $\overline{E}_n$ corresponding to each possible intermediate state $|n\rangle$ that can be inserted in the hadronic amplitude $\left\langle 0\left|T\left\{ J_\mu(u) J_\nu(v) \mathcal{H}_W(0)\right\}\right|K_L\right\rangle$ is larger than $M_K$.  Here the added bar on the energy $\overline{E}_n$ indicates an intermediate-state energy that is the sum of the energy $E_n$ of the hadronic state $|n\rangle$ and the energy of the photon or photons that are created at the locations $u$ and/or $v$ of the two currents $J_\mu(u) J_\nu(v)$  in this amplitude.  

A simple Wick rotation is not possible for the $p_0$ integral in Eq.~\eqref{eq:K-MS} because of the $p_0$ singularities present in the two photon propagators in that equation.  In fact, we must preserve the Minkowski-space character of this $p_0$ integral if we are to retain the imaginary part of $\mathcal{A}_{ss'}(k^+,k^-)$ required by Cutkowski's rules and the real part that contains the physically-required principal part prescription of the integral over the  two-photon energy, $E_{\gamma\gamma}$.  This principal part prescription is needed to define the $E_{\gamma\gamma}$ integral in the presence of the singularity when $E_{\gamma\gamma} = M_K$.  As in the case of the $\pi^0\to e^+ e^-$ decay it is possible to Wick rotate the $u_0$ and $v_0$ contours while retaining the appropriate Minkowski character of the $p_0$ integral.

We will establish this by considering three cases.  We begin by assuming $u_0 \ge v_0$ and identify the three cases as A) $u_0 \ge v_0 \ge 0$, B) $u_0 \ge 0 \ge v_0$ and C) $0 \ge u_0 \ge v_0$.  The pairs of variables $(u, \mu)$ and $(v, \nu)$ enter the hadronic amplitude on the second line of Eq.~\eqref{eq:K-MS} symmetrically.  While these pairs do not enter the electromagnetic factor appearing in the first and second lines of Eq.~\eqref{eq:K-MS} symmetrically, their exchange is related by charge conjugation symmetry so that the singularity structure is not affected by their exchange, justifying the restriction to $u_0 \ge v_0$.

\subsection{Case A}
\label{sec:case-A}

\begin{figure}[h]
\centering
\includegraphics[width=0.5\textwidth]{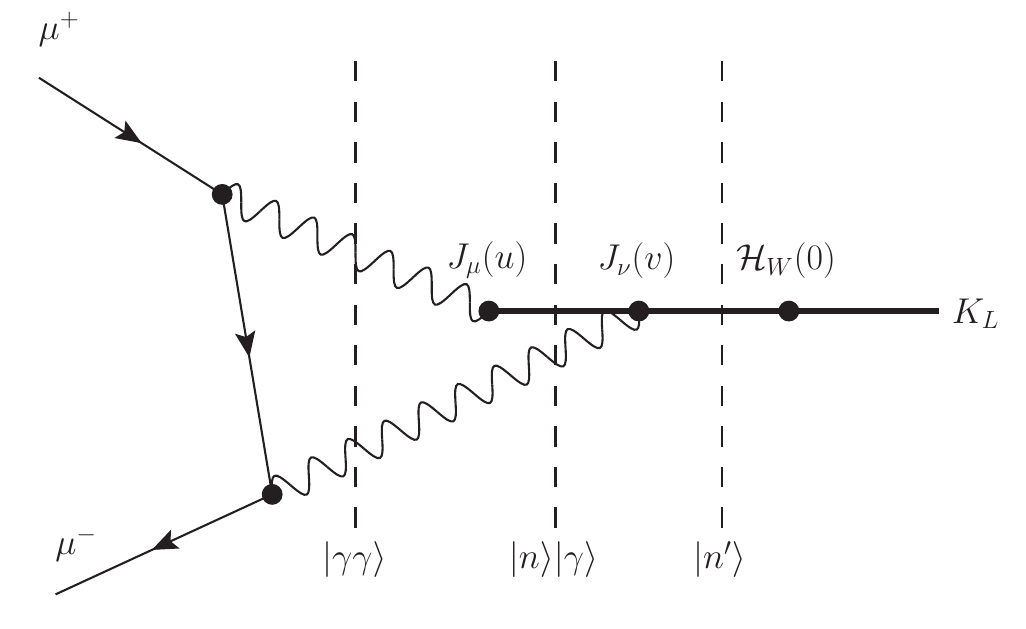}
\caption{The diagram representing the amplitude in case A that appears in Eq.~\eqref{eq:K-MS-A}.  The time ordering of the three hadronic operators is indicated by their horizontal positions and the three intermediate states of importance are also labeled. The earlier time is on the right of the plot and ascends as one goes to the left.}
\label{fig:KLmumu-A}
\end{figure}

Begin with case $u_0 \ge v_0 \ge 0$ and consider the integration over only the time-components, $u_0$, $v_0$ and $p_0$ for a particular pair of hadronic intermediate states $|n\rangle$ and $|n'\rangle$ with energies $E_n$ and $E_{n'}$.  Thus, we need to study the Minkowski-space expression:
\begin{eqnarray}
\int_0^T\!\! d u_0 \; \int_0^{u_0}\!\! d v_0 \; \int_{-\infty}^\infty\!\! d p_0\;\frac{e^{i\left(\frac{M_K}{2}+p_0\right)u_0}  e^{i\left(\frac{M_K}{2}-p_0\right)v_0}}{(\frac{P}{2}+p)^2+m_\gamma^2-i\varepsilon} \frac{e^{-i E_n u_0} e^{-i (E_{n'} - E_n)v_0}}{(\frac{P}{2}-p)^2+m_\gamma^2-i\varepsilon}
\label{eq:K-MS-A} && \\
&& \hskip -4.6 in \times
   \frac{\overline{u}_s(k^-)\gamma_\nu\{\gamma\cdot(\frac{P}{2}+p-k^+) +m_\mu\}\gamma_\mu v_{s'}}
      {(\frac{P}{2}+p-k^+)^2 +m_\mu^2-i\varepsilon} \nonumber 
      \left\langle 0\left|J_\mu(0)|n\rangle\langle n| J_\nu(0)|n'\rangle\langle n'| \mathcal{H}_W(0)\right|K_L\right\rangle, 
\nonumber
\end{eqnarray}
where we have introduced an upper limit $T$ to the $u_0$ integration so that the limit $T\to\infty$ must be evaluated. The diagram corresponding to the amplitude appearing in Eq.~\eqref{eq:K-MS-A} is shown in Fig.~\ref{fig:KLmumu-A} where the horizontal ordering of the three hadronic operators represents their time ordering and the intermediate states cut by the vertical dotted lines are labeled.   

If we integrate over $v_0$ we are left with the integration over $p_0$ and $u_0$ with the following exponential factors appearing in the integrand:
\begin{equation}
\int_0^T du_0 \int_{-\infty}^\infty dp_0 \left\{e^{i(M_K-E_{n'})u_0} - e^{i(\frac{M_K}{2} + p_0 -E_n)u_0}\right\}.
\label{eq:case-A}
\end{equation}
The left-hand term in this expression corresponds to the $|n'\rangle$ intermediate state, identified by the right-most dotted line in Fig.~\ref{fig:KLmumu-A}, and permits a simple Wick rotation of the $u_0$ contour from the positive real axis to the negative imaginary axis provided $E_{n'} > M_K$.  This inequality will be obeyed provided we have removed and evaluate separately the contributions from the single-pion state.  
(An intermediate vacuum  or two-pion states are not possible because we are considering the decay of the CP-odd $K_L$ meson and neglecting CP-violating effects.)
As in the calculation of the $K_L-K_S$ mass difference, the small difference between the kaon and $\eta$ mass requires that the $\eta$ state also be removed and the $\eta$ state also evaluated separately.
Assuming that these states have been removed from the hadronic matrix elements, we can apply Cauchy's theorem to distort the $u_0$ contour from the positive real axis to the negative imaginary axis.  When doing so, there will also be a quarter-circle integration contour that extends from the $+T$ end point of the original real contour to the $-iT$ end point of the desired imaginary contour.  Because of the sign of $E_{n'}-M_K$, the contribution from this unwanted contour will decrease exponentially as $T$ increases so that the Minkowski and Euclidean $u_0$ integrals become identical.

\begin{figure}[h]
\centering
\includegraphics[width=0.65\textwidth]{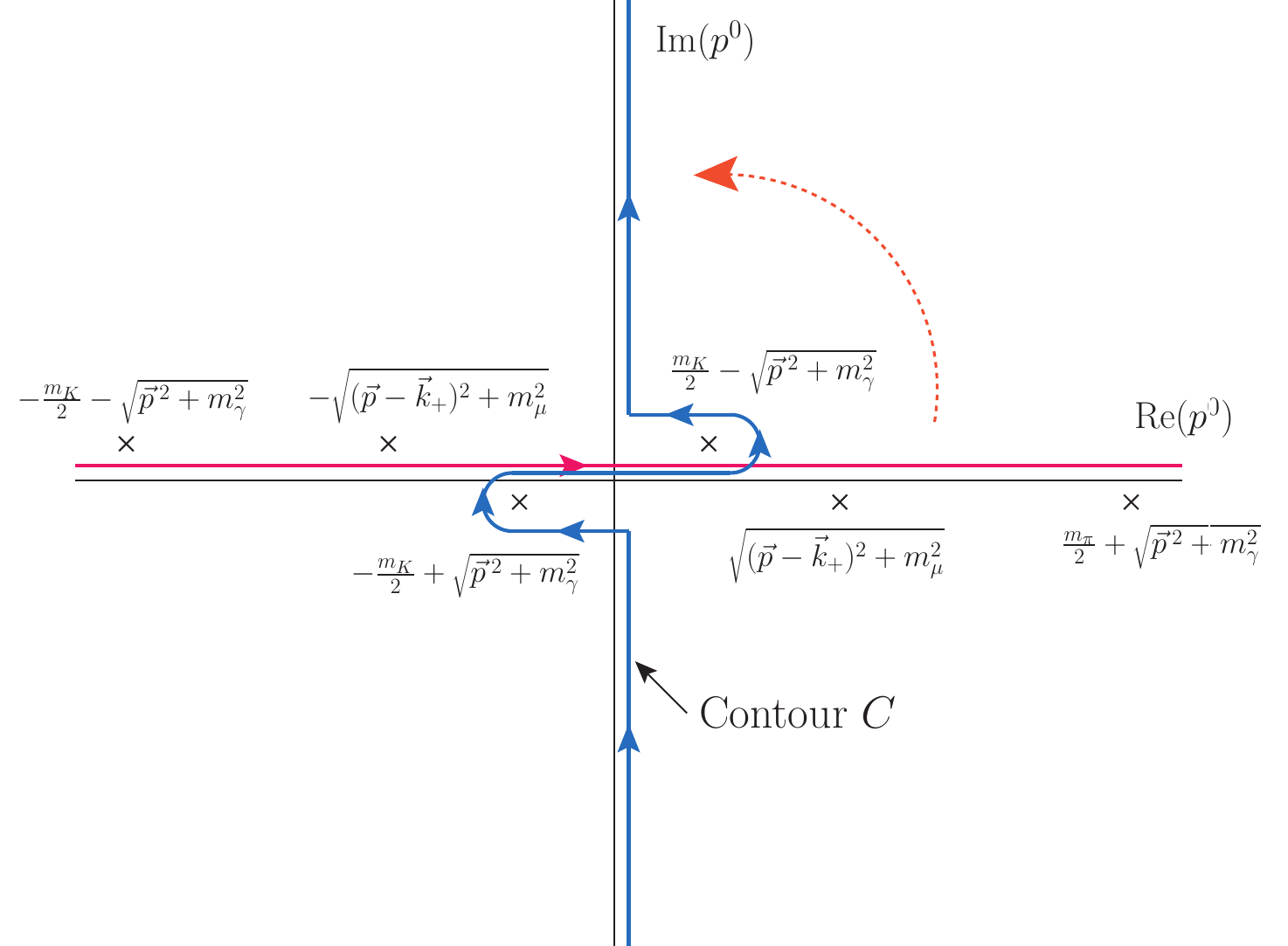}
\caption{The original (red) deformed (blue) $p_0$ integration contours. The six crosses locate the six poles that arise from the three propagators which appear in the integrand in Eq.~\eqref{eq:K-MS}.  The conclusions drawn in this section required that the original contour for $p_0$ real and positive be deformed as shown in this figure so that Re$(p_0)$ remains less than $\frac{m_K}{2}$.  There were no lower bounds needed for Re$(p_0)$ so the symmetrical deformation of the negative real part of the $p_0$ contour shown here is not required.}
\label{fig:p0-contour}
\end{figure}

The right-hand term in Eq.~\eqref{eq:case-A} requires that the $p_0$ integration, which is to be carried out before the $u_0$ integration, must  also be considered.  The original $p_0$ contour lies along the real axis and is shown in Fig.~\ref{fig:p0-contour}.  Since we intend to again rotate the $u_0$ integration from the positive real to the negative imaginary axis, we must show that the $p_0$ integration contour can be deformed so that $E_n-\frac{M_K}{2}-\mathrm{Re}(p_0) \ge 0$ for all values of $p_0$.  Examining the location of the singularities in the complex $p_0$-plane shown in Fig.~\ref{fig:p0-contour}, we see that the real part of $p_0$ can be bounded from above by $\frac{M_K}{2}-\sqrt{\textbf{p}^2 + m_\gamma^2} \ge \mathrm{Re}(p_0)$ if we rotate the $p_0 \ge 0$ part of the original $p_0$-contour to follow the positive imaginary axis as shown in Fig.~\ref{fig:p0-contour}.  In this case, as in the case of the $u_0$ contour discussed above, the quarter-circle ``contour at infinity'' joining the original and deformed $p_0$-contours is exponentially suppressed.  This suppression results because with $u_0$ initially real and positive the $p_0$-dependent exponent appearing in the right-hand term in Eq.~\eqref{eq:case-A}, $+ip_0 u_0$, acquires a negative real part in the first quadrant of the complex $p_0$ plane. (Note, even as $u_0\to 0$, the convergence of the $p_0$ integral is guaranteed by the $p_0$ dependence of the muon and photon propagators.)  Thus, if $E_n$ obeys the following condition, then the desired Wick rotation of the $u_0$ integral is allowed:
\begin{eqnarray}
E_n-\frac{M_K}{2}-\mathrm{Re}(p_0) \ge E_n - M_K + \sqrt{\textbf{p}^2 + m_\gamma^2} \ge 0
   \quad\mbox{or}\quad 
E_n + \sqrt{\textbf{p}^2 + m_\gamma^2} \ge M_K.
\label{eq:bound-A}
\end{eqnarray}
We conclude that the needed Wick rotation of the $u_0$ contour will be possible if the energy of the $|n\rangle|\gamma(\textbf{p})\rangle$ intermediate state is larger than $M_K$, the expected constraint.  

The hadronic state $|n\rangle$ with the smallest energy that can appear in such an intermediate state is the two-pion state.  In this case $E_n + \sqrt{\textbf{p}^2 + m_\gamma^2} \ge 2 m_\pi + m_\gamma$, violating the bound given in Eq.~\eqref{eq:bound-A}.  In Section~\ref{sec:pi-pi-gamma} we will argue that the contribution of a $\pi\pi\gamma$ intermediate with energy at or below $M_K$ will be highly suppressed by three-body phase space and give an estimate for a plausible accuracy target in a $K_L\to\mu^+\mu^-$ calculation where this state can be safely neglected.

\begin{figure}[h]
\centering
\includegraphics[width=0.5\textwidth]{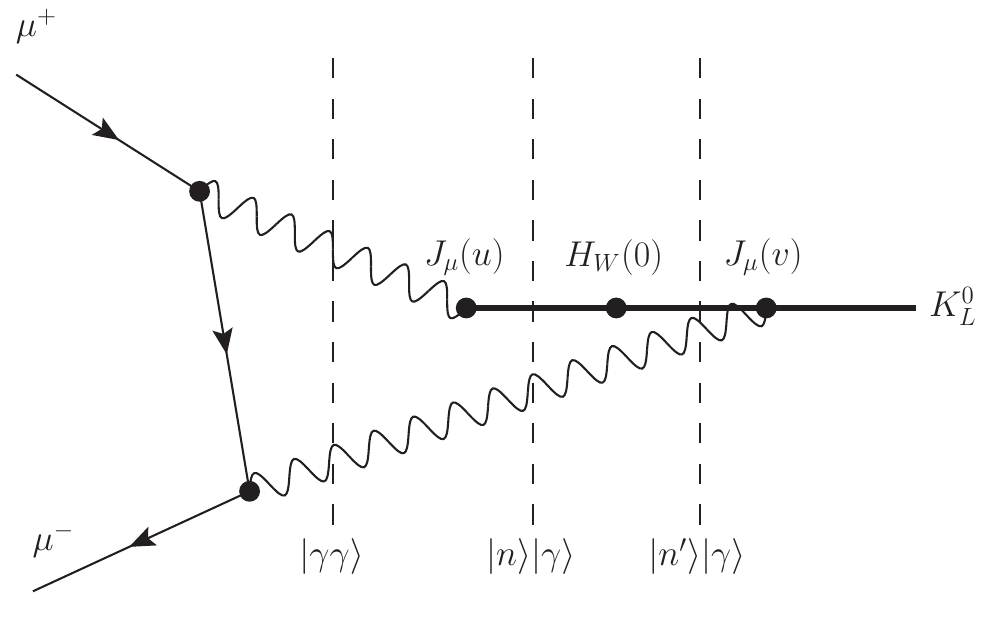}
\caption{The diagram which corresponds to the case B time ordering which appears in Eq.~\eqref{eq:K-MS-B}.  This time ordering of the three hadronic operators is indicated by their horizontal positions and the three intermediate states of importance are also labeled.}
\label{fig:KLmumu-B}
\end{figure}

\subsection{Case B}
\label{sec:case-B}

Next we consider the time ordering $u_0 \ge 0 \ge v_0$ represented schematically in Fig.~\ref{fig:KLmumu-B}.  In this case the exponential factors and operator ordering that appear for case A in Eq.~\eqref{eq:K-MS-A} take the form:
\begin{eqnarray}
e^{i\left(\frac{M_K}{2}+p_0\right)u_0}  e^{i\left(\frac{M_K}{2}-p_0\right)v_0}  e^{-i E_n u_0} e^{-i (M_K-E_{n'})v_0}\left\langle 0\left|J_\mu(0)|n\rangle\langle n| \mathcal{H}_W(0) |n'\rangle\langle n'|J_\nu(0)\right|K_L\right\rangle, 
\label{eq:K-MS-B}
\end{eqnarray}
We next we add the integrations over $u_0$, $v_0$ and $p_0$ to the expression in Eq.~\eqref{eq:K-MS-B} and drop the now time-independent hadronic matrix element to obtain:
\begin{equation}
\int_0^T du_0 \int_{-T}^0 d v_0 \int_{-\infty}^\infty dp_0 \left\{e^{i(\frac{M_K}{2}+p_0-E_{n})u_0} 
    e^{-i(\frac{M_K}{2} + p_0 -{E_n'})v_0}\right\}.
\label{eq:case-B}
\end{equation}

The integrals that appear in Eq.~\eqref{eq:case-B} are very similar to those just discussed for the right-hand term in Eq.~\eqref{eq:case-A}.  In the case of Eq.~\eqref{eq:case-B} the differing signs of the exponents in the two exponential factors are compensated by the different signs of the $u_0$ and $v_0$ variables.   Thus, as in that earlier case, both of the exponential factors allow us to deform the $p_0$-contour that follows the positive real axis to instead follow the blue contour along the positive imaginary axis in Fig.~\ref{fig:p0-contour} while leaving the $p_0$ contour along the negative real axis unchanged.  We can then Wick rotate the $u_0$ contour specified in Eq.~\eqref{eq:case-B} that lies on the positive real axis to the negative imaginary axis and the $v_0$ contour along the negative real axis to follow the positive imaginary axis.  The first will be possible if $E_n-\frac{M_K}{2}-\mathrm{Re}(p_0) \ge 0$ while the second deformation can be performed if $E_{n'}-\frac{M_K}{2}-\mathrm{Re}(p_0) \ge 0$.

As in case A, the new $p_0$ contour adopted above obeys Re$(p_0) \le \frac{M_K}{2} - \sqrt{\textbf{p}^2 + m_\gamma^2}$ so that the two conditions in the previous paragraph will be obeyed if:
\begin{eqnarray}
E_X + \sqrt{\textbf{p}^2 + m_\gamma^2} \ge M_K.
\label{eq:bound-B}
\end{eqnarray}
for $X=n$ and $X=n'$.  Again the Wick rotations of the $u_0$ and $v_0$ contours needed if the hadronic matrix element is to be evaluated in Euclidean space are possible if the energies of the two right-most intermediate states identified in Fig.~\ref{fig:KLmumu-B} are larger than $M_K$.  Since the state $|n'\rangle$ must necessarily carry strangeness, Eq.~\eqref{eq:bound-B} will be obeyed for the case $X = n'$.   As in case A, only when the state $|n\rangle$ is a two-pion state will the desired Wick rotation of the $u_0$ contour be impossible.  As in case A, we will assume that this contribution is sufficiently small that it can be neglected.

\begin{figure}[h]
\centering
\includegraphics[width=0.5\textwidth]{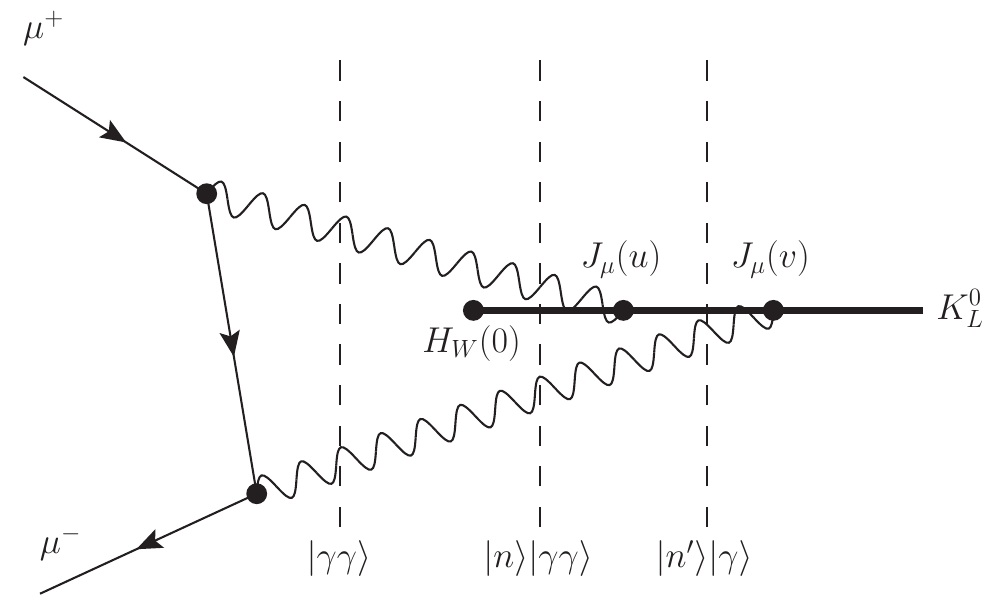}
\caption{The diagram which corresponds to the time ordering of case C which appears in Eq.~\eqref{eq:K-MS-C}.  This time ordering of the three hadronic operators is indicated by their horizontal positions and the three intermediate states of importance are also labeled.}
\label{fig:KLmumu-C}
\end{figure}

\subsection{Case C}
\label{sec:case-C}

Finally we consider the time order $0 \ge u_0 \ge v_0$.  This ordering is described by the diagram in Fig.~\ref{fig:KLmumu-C} and the critical portion of the amplitude is shown in the expression:
\begin{eqnarray}
\hskip -0.2 in 
e^{i\left(\frac{M_K}{2}+p_0\right)u_0}  e^{i\left(\frac{M_K}{2}-p_0\right)v_0}  e^{-i (E_{n'}-E_n) u_0} e^{-i (M_K-E_{n'})v_0}  &&
   \label{eq:K-MS-C} \\ 
 &&\hskip -0.5 in \times \left\langle 0\left|\mathcal{H}_W(0)|n\rangle\langle n|J_\mu(0) |n'\rangle\langle n'|J_\nu(0)\right|K_L\right\rangle\!\!, 
\nonumber
\end{eqnarray}
analogous to Eq.~\eqref{eq:K-MS-B} for case B.  Including the integration over $u_0$, $v_0$ and $p_0$ and dropping the time-independent hadronic matrix element, we arrive at an expression analogous to that in Eq.~\eqref{eq:case-B} for case B:
\begin{equation}
\int_{-T}^0 dv_0 \int_{v_0}^0 d u_0 \int_{-\infty}^\infty dp_0 \left\{e^{i(\frac{M_K}{2}+p_0+E_n-E_{n'})u_0}                                      e^{-i(\frac{M_K}{2} + p_0 -{E_{n'}})v_0}\right\}.
\label{eq:case-C1}
\end{equation}
If we next perform the integral over $u_0$ the result takes the form:
\begin{equation}
\int_{-T}^0 dv_0 \int_{-\infty}^\infty dp_0 \left\{e^{i(E_{n'} -\frac{M_K}{2} - p_0)v_0} - e^{iE_n v_0} \right\}.
\label{eq:case-C2}
\end{equation}
The right-hand term within the curly brackets presents no difficulty since $E_n$ is positive and the $v_0$ contour can be rotated from the negative real axis to the positive imaginary axis using Cauchy's theorem.  The contribution of the quarter-circle contour joining the end point $-T$ of the original contour and the end point $iT$ of the rotated contour falls exponentially with increasing $T$ and therefore does not contribute.  (That this term evaluated at $u_0=v_0$ should depend only on the distance between $u_0=v_0$ and the location of $\mathcal{H}_W(0)$ and should fall exponentially in Euclidean space as $\exp\{-E_n|v_0|\}$ for increasing $|v_0|$ can be anticipated from Fig.~\ref{fig:KLmumu-C}.)

The left-hand term within the curly brackets requires more attention since a similar Wick rotation of the $v_0$ contour is possible if $E_{n'} -\frac{M_K}{2} - \mathrm{Re}(p_0) \ge 0$.  This condition can be met if the portion of the original $p_0$ which follows the positive real axis can be deformed as shown in Fig.~\ref{fig:p0-contour} so that Re$(p_0) \le \frac{M_K}{2} -\sqrt{\textbf{p}^2+m_\gamma^2}$.  The real negative value of $v_0$ makes such a contour rotation possible.  Thus, a sufficient condition permitting the Wick rotation of the $v_0$ contour becomes
\begin{equation}
E_{n'} -\frac{M_K}{2} \ge \frac{M_K}{2} -\sqrt{\textbf{p}^2+m_\gamma^2} \quad \mbox{or} \quad
E_{n'} + \sqrt{\textbf{p}^2+m_\gamma^2} \ge M_K,
\label{eq:bound-C}
\end{equation}
the expected result and a condition that is clearly obeyed for the case at hand since the state $|n'\rangle$ carries strangeness.

\subsection{Summary formula}

Here we give the result that is obtained by following the steps described in the previous three sections.  We first remove the contributions from intermediate QCD energy eigenstates $|n\rangle$ or QCD and single-photon energy eigenstates $|n\rangle|\gamma\rangle$ states whose total energy is less than the kaon mass.  This permits the rotation of the $u_0$ and $v_0$ contours so that the amplitude given in Eq.~\eqref{eq:K-MS} can be written as a position-space integral of the product of a complex leptonic kernel function $L_{\mu\nu}$ and a Euclidean-space hadronic matrix element:
\begin{eqnarray}
    \mathcal{A}^{\mathrm{I}} &=& \int_{-{T^-_v}}^{T^+_v} dv_0\int_V d^3\textbf{v}
  \int_{v_0}^{T_u+v_0} du_0\int_V d^3\textbf{u}\; e^{M_K(u_0+v_0)/2}
  \label{eq:Result-conv}\\
  && \hskip 2.4 in L_{\mu\nu}(u-v) 
  \langle T\left\{ J_\mu(u)J_\nu(v) \chw(0) K_L(t_i)\right\} \rangle'.
  \nonumber
\end{eqnarray}
Here we have limited the space-time integrals appearing in Eq.~\eqref{eq:K-MS} to correspond to that region which must be included in a lattice QCD calculation to insure exponential accuracy in the limit that this restricted volume becomes large.  Specifically, the spatial integrals have been limited to the finite volume $V$ of the lattice calculation (with periodic boundary conditions) and the time limits $T_v^\pm$ and $T_u$ must be chosen sufficiently large to include the full two-dimensional temporal region around the location of the weak Hamiltonian in which the hadronic four-point function is non-zero. Finally, to be concrete, we have introduced a normalized kaon interpolating operator $K_L(t_i)$ located at the initial time $t_i$ which creates the initial kaon.  The initial time $t_i$ must be chosen sufficiently more negative than the $-T^-_v$ that only the kaon state is able propagate between $t_i$ and the nearest current operator $J_\nu(\textbf{v},-T_v^-)$.  The superscript $\mathrm{I}$ on $\mathcal{A}^{\mathrm{I}}$ indicates both these changes making the amplitude accessible to lattice QCD and the omission of those energy eigenstates $|n\rangle$ and $|n\rangle|\gamma\rangle$ with a total energy that is smaller than $M_K$.  The prime on the hadronic matrix element in Eq.~\eqref{eq:Result-conv} is introduced to indicate that the contributions of these low energy $|n\rangle$ and $|n\rangle|\gamma\rangle$ states have been explicitly subtracted.

In order to obtain the complete $K_L\to\mu\mu$ decay amplitude we must also include the physical contribution of those low-energy states whose entire contribution has been removed from Eq.~\eqref{eq:Result-conv}. The only QCD-QED product states $|n\rangle|\gamma\rangle$ with total energy below $M_K$ are the $\pi\pi\gamma$ states whose effects we plan to neglect.  These are discussed in the next section and will not be included here.  Thus, we will deal only with the QCD energy eigenstates $|n\rangle$ with energy $E_n \le M_K$.  Only the single-pion state $\pi^0$ meets this criteria. However, as mentioned above we also plan to make a similar accommodation for the $\eta$ state.  For the case of the $\eta$ its unphysical contribution decreases so slowly as the integration range $T^+_v$ is increased that it is best also to be explicitly removed.

Those low-energy QCD eigenstates $|n\rangle$ which require special treatment can appear as intermediate states only between the current $J_\nu(v)$ and the weak Hamiltonian $\chw(0)$.  Given our choice of limits in Eq.~\eqref{eq:Result-conv}, we can separate the integrals over $v_0$ and $u_0$ if we replace the $u_0$ variable by $w_0 = u_0-v_0$.  If in addition we assume $v_0\ge 0$ and insert the intermediate state $|n\rangle$ between the operators $J_\nu(v)$ and $H_W(0)$, we can then perform the integration over $v_0$ to obtain:
\begin{eqnarray}
    \mathcal{A}^n &=& \int_V d^3\textbf{v}
  \int_{0}^{T_u} dw_0\int_V d^3\textbf{u}\; e^{M_K w_0/2} L_{\mu\nu}(\textbf{u}-\textbf{v},w_0)
  \label{eq:Result-growing}\\
  && \hskip 1.0 in  \times\left[\frac{e^{(M_K-E_n)T^+_v}-1}{M_K-E_n}\right]
  \langle T\left\{ J_\mu(\textbf{u},w_0)J_\nu(\textbf{v},0)\right\}|n\rangle\langle n|T\left\{\{\chw(0) K_L(t_i)\right\} \rangle.
  \nonumber
\end{eqnarray}
While it is this entire amplitude which has been removed from $\mathcal{A}^{\mathrm{I}}$ given in Eq.~\eqref{eq:Result-conv}, it is only the $e^{(M_K-E_n)T^+_v}$ term in the square brackets in Eq.~\eqref{eq:Result-growing} which is unphysical.  The $-1$ term in the square brackets is part of the physical amplitude and must be restored.  Thus in addition to amplitude $\mathcal{A}^{\mathrm{I}}$ given in Eq.~\eqref{eq:Result-conv} we must also compute and include a second amplitude 
\begin{eqnarray}
    \mathcal{A}^{\mathrm{II}} &=& -\sum_n\int_V d^3\textbf{v}
  \int_{0}^{T_u} dw_0\int_V d^3\textbf{u}\; [\frac{e^{M_K w_0/2}}{M_K-E_n}L_{\mu\nu}(\textbf{u}-\textbf{v},w_0)
  \label{eq:Result-growing}\\
  && \hskip 1.5 in  
  \times \langle T\left\{ J_\mu(\textbf{u},w_0)J_\nu(\textbf{v},0)\right\}|n\rangle
  \langle n|T\left\{\chw(0) K_L(t_i)\right\} \rangle,
  \nonumber
\end{eqnarray}
where the sum is performed over all states $|n\rangle$ that were removed from the amplitude $\mathcal{A}^{\mathrm{I}}$ given in Eq.~\eqref{eq:Result-conv}.  Thus, the complete amplitude $\mathcal{A}_{K_L\mu\mu}$ is given by the sum of $\mathcal{A}^{\mathrm{I}}$ and $\mathcal{A}^{\mathrm{II}}$:
\begin{equation}
\mathcal{A}_{K_L\mu\mu} = \mathcal{A}^{\mathrm{I}} + \mathcal{A}^{\mathrm{II}}.
\label{eq:total_physical}
\end{equation}

For completeness we also include here a formula for the kernel $L_{\mu\nu}$, restricted to the case of spatial indices -- the case needed when the kaon is at rest:

\begin{eqnarray}
  L_{ij}(w) &=& 2  m_\mu \alpha^2 \epsilon_{ijk}\frac{w^k }{|\mathbf{w}|^2} \left[- \frac{e^{\frac{M_K}{2} |w_0| }}{M_K |\mathbf{k}_+|} \ln \left(\frac{1 + \beta}{1 - \beta}\right) \int_0^\infty d |\mathbf{p}| \frac{e^{-|\mathbf{p}||w_0|}}{M_K - 2|\mathbf{p}|+i\varepsilon} F(|\mathbf{p}||\mathbf{w}|) \right. \nonumber \\
  && \hskip -0.8 in 
            + \; \frac{e^{-\frac{M_K}{2} |w_0| }}{M_K |\mathbf{k}_+|} \ln \left(\frac{1 + \beta}{1 - \beta}\right) \int_0^\infty d|\mathbf{p}| \frac{e^{-|\mathbf{p}||w_0|}}{M_K + 2|\mathbf{p}|} F(|\mathbf{p}||\mathbf{w}|)  \\
            && \hskip -0.8 in + \left.  2\int_0^\infty\!\!  d|\mathbf{p}|\int_{-1}^1  d\cos\theta\; \frac{e^{-E_\mu(|\mathbf{p}|,|\mathbf{k}_+|,\theta)\;|w_0|}}{E_\mu(|\mathbf{p}|,|\mathbf{k}_+|,\theta)(-M_K + 2|\mathbf{k}_+|\cos\theta)(M_K + 2|\mathbf{k}_+|\cos\theta)} F(|\mathbf{p}||\mathbf{w}|) \right]\,,\nonumber  \label{eq:Luv}
\end{eqnarray}
where $E_\mu(|\mathbf{p}|,|\mathbf{k}_+|,\theta) =\sqrt{|\mathbf{p}|^2 + |\mathbf{k}_+|^2 - 2|\mathbf{p}||\mathbf{k}_+|\cos\theta + m_\mu^2}\;$ is the energy of the intermediate muon and $
F(t)= \cos(t)
- \frac{1}{t}\sin(t)$.

\section{Estimate of low-energy $\pi\pi\gamma$ contribution}\label{sec:pi-pi-gamma}

As explained in Sect.~\ref{sec:eucl}, the existence of low-energy $\pi\pi\gamma$ intermediate states is one of the uncontrolled difficulties within our proposed computational framework. We will refer to $\pi\pi\gamma$ states in the kinematic region where $E_{\pi\pi} + |\textbf{p}| \le M_K$ as propagating $\pi\pi\gamma$ states because they propagate in Euclidean time with less suppression with increasing time than the kaon and, as a result, the discrete finite-volume $\pi\pi$ eigenstates involve spatial propagation that creates sensitivity to the finite-volume of the QCD calculation.  (Here $E_{\pi\pi}$ is the energy of the $\pi\pi$ state and $\textbf{p}$ the three momentum of the photon, both in the kaon rest system.)

These propagating $\pi\pi\gamma$ states introduce two potential problems.  First their contribution will grow exponentially as the separation between the source and sink operators in the lattice calculation increases.  Second, even if such unphysical, exponentially growing terms could be removed there is no known procedure to restore the physical contribution that should have resulted from a proper treatment of these propagating $\pi\pi\gamma$ states.  In this section we will discuss each of these difficulties in turn.  In both cases we will argue that these unwanted effects are small and should be expected to contribute below the 5\% level for lattice volumes with a linear size of at least 10 fm or smaller.

\subsection{Exponential growth from propagating $\pi\pi\gamma$ states}

We first consider the unphysical contribution of the propagating $\pi\pi\gamma$ states to our Euclidean calculation.  To understand their effects we need only consider the finite-volume two-pion energy eigenstates with discrete energies $E_{\pi\pi} < M_K$.  Specifically we consider the time ordering of case A with $u_0 \ge v_0 \ge 0$ (see Sec.~\ref{sec:case-A}), insert a complete set of QCD energy eigenstates between $J_\mu(u)$ and $J_\nu(v)$, and then focus on the two-pion states within that sum.  The energy of the combined $\pi\pi\gamma$ intermediate state is then $E_{\pi\pi}+|\textbf{p}|$, where $\textbf{p}$ is the three momentum of the photon emitted by the $J_\nu(v)$.  

In our QED$_\infty$ treatment of electromagnetism the additional energy of the photon carrying momentum $\textbf{p}$ can be arbitrarily small.  One might naively expect that momentum conservation would constrain the photon momentum $\textbf{p}$ to those discrete values equal to the negative of the discrete values allowed for the total momentum carried by the recoiling $\pi\pi$ state.  However, this conclusion is not correct for QED$_\infty$ since the photon and hadron degrees of freedom move in overlapping but different volumes and have different momentum spectra with very different eigenvectors that obey no mutual orthogonality conditions.  

The allowed total momentum $\textbf{p}_{\pi\pi}$ of the $\pi\pi$ QCD eigenstates takes the usual finite-volume form $2\pi(n_1,n_2,n_3)/L$, determined by the three integers $\{n_i\}_{1 \le i \le 3}$ while the photon momentum $\textbf{p}$ is a 3-vector with continuous components.  Both the position dependence of the  infinite-volume photon wave function with momentum $\textbf{p}$ and the dependence on the center-of-mass position of the finite volume $\pi\pi$ wave function are known exactly.  This allows us to evaluate explicitly both the integrals over $\textbf{u}$ and $\textbf{v}$ in Eq.~\eqref{eq:Result-conv} to obtain the unfamiliar form of such a combination of finite- and infinite-volume quantities.  The resulting inner products between these different total-momentum eigenstates contribute an explicit factor of
\begin{equation}
\frac{1}{(2\pi L)^3}\prod_{k=1,2,3} \left[\frac{\sin\left(\pi n_k +\frac{p_k}{2} L \right)}
{\frac{\pi n_k}{L} +\frac{p_k}{2}}\right]^2
\label{eq:FV-delta}
\end{equation}
to the $K\to\mu^+\mu^-$ decay amplitude.  This factor correlates but does not equate the momenta $\textbf{p}$ of the photon and the negative of the total momentum $\textbf{p}_{\pi\pi}$ of the $\pi\pi$ state. In the limit $L\to\infty$ the expression in Eq.~\eqref{eq:FV-delta} approaches a simple three-dimensional delta function evaluated at the sum of the momenta $\textbf{p}_{\pi\pi}$ and $\textbf{p}$.

The other important factors in the contribution of these $\pi\pi\gamma$ intermediate states are the actual matrix elements of the two currents:
\begin{equation}
\langle 0| J_\mu(0)|\pi\pi(\textbf{p}_{\pi\pi})\rangle
\langle\pi\pi(\textbf{p}_{\pi\pi})|J_\nu(0)H_{\rm W}|K_L\rangle.
\label{eq:pipi-FV}
\end{equation}
As discussed in the next section, the size of these factors estimated from the known $K_L\to\pi\pi\gamma$ decay rate leads to a contribution to the $K_L\to\mu^+\mu^-$ decay amplitude at the percent level.  Here we are concerned with the possible amplification of this contribution coming from the exponential growth at large time separations created by intermediate $\pi\pi\gamma$ states with energy below $M_K$.  To simplify the estimation of these exponentially growing terms we will neglect $\pi\pi$ interactions and treat the finite-volume $\pi\pi$ states as composed of free particles obeying periodic boundary conditions.  With this assumption, the factor $\langle\pi\pi(\textbf{p}_{\pi\pi})|J_\nu(0)H_{\rm W}|K_L\rangle$ in Eq.~\eqref{eq:pipi-FV} must have the form given in Eq.~\eqref{eq:vpipig}  in the section below. This requires that both pions carry non-zero momenta: $\textbf{p}^+$ and $\textbf{p}^-$.

If the exponential growth is associated with a momentum conserving process with $\textbf{p} = -\textbf{p}_{\pi\pi} = -2\pi \textbf{n}/L$ so that the factor in Eq.~\eqref{eq:FV-delta} is not suppressed by inverse powers of $L$, then we would require
\begin{equation}
2\sqrt{M_\pi^2 + \left(\frac{2\pi}{L}\right)^2} + \sqrt{2}\frac{2\pi}{L} < M_K\,,
\label{eq:on-shell-expn-grow}
\end{equation}
where we have assumed that the two pions carry minimum but distinct lattice momenta with $|\textbf{p}^\pm| = 2\pi/L$.  The inequality in Eq.~\eqref{eq:on-shell-expn-grow} requires $L > 11.8$ fm.

As indicated by Eq.~\eqref{eq:FV-delta}, momentum conservation is not required so a smaller photon momentum $\textbf{p}$ than that used in Eq.~\eqref{eq:on-shell-expn-grow} is possible, decreasing the minimum energy of a propagating $\pi\pi\gamma$ intermediate state.
For $L \ge 6$ fm, $E_{\mathrm{min}}=2\sqrt{(M_\pi)^2 + (\frac{2\pi}{L})^2}$ will fall below $M_K$ and this non-conservation of momentum will allow an exponentially growing $\pi\pi\gamma$ state.
However, the small phase space associated with the limited photon momentum and the multiple powers of $1/L$ resulting from both the $L$ dependence of the factor given in Eq.~\eqref{eq:FV-delta} and as well as additional inverse powers of $L$ that will come from the integration over continuous components of $\textbf{p}$ are expected to make such terms negligible.

We should acknowledge that the presence of these propagating $\pi\pi\gamma$ states for $L \ge 6$ fm, is a result of using QED$_\infty$.  In many applications this formulation of QED removes power-law finite volume errors, giving a potentially significant increase in the accuracy of the QED results.  If these propagating $\pi\pi\gamma$ states were not present in this process, for example if the pion mass were larger than $M_K/2$, then this would also be true for the process at hand.  The Euclidean-space result for the product of $H_W$ and the two electromagnetic currents would vanish exponentially as the space-time separations between these three operators is increased.  By restricting the calculation to a volume of side $L$ obeying periodic boundary conditions we would be able to capture this non-zero region with errors that fall exponentially with increasing $L$.  

This favorable situation is degraded by the propagating $\pi\pi\gamma$ states.  The two pions in these low-energy states can propagate without exponential suppression through out the volume in which the lattice QCD calculation is performed, introducing an unsuppressed dependence on that volume.  Here we have argued that the presence of these states has a relatively small effect on the calculation proposed here so these effects can be safely neglected.  By using QED$_\infty$ we insure that the contributions of the more important, higher-energy photons do not introduce power law finite volume errors as would be the case were we to use, for example, the QED$_{\mathrm{L}}$ finite-volume formulation of QED~\cite{Hayakawa:2008an}.

\subsection{Importance of propagating $\pi\pi\gamma$ states in the physical $K_L\to\mu^+\mu^-$ decay}

In the previous section we discussed the contribution of the propagating $\pi\pi\gamma$ states to a lattice calculation of the $K_L\to\mu^+\mu^-$ decay amplitude based on QED$_\infty$.  Anticipating the discussion in this section, we stated that the contribution of these states to the physical decay is small and that the Euclidean complication of exponential growth would be smaller still for the typical lattice volumes of interest.  Even if we were able to identify and remove these unphysical, exponentially growing terms from a lattice QCD calculation, the remaining ``physical'' contributions are expected to have large finite volume errors and, in contrast to the contribution of propagating $\pi\pi$ states in the calculation of $\Delta M_K$~\cite{Christ:2015pwa}, there is at present no known method to remove those finite volume errors.  

Fortunately, both the finite-volume-distorted contribution of these propagating states to our lattice calculation and the physical contribution of these states to the actual decay (contributions not provided by a lattice calculation) are small and can be safely neglected in a calculation with a target accuracy of 10\%.  In this section we will address the physical decay and demonstrate that the contribution from the $\pi\pi\gamma$ states in this low-energy, ``propagating'' region is at the few percent level.

\begin{figure}[h!]
\includegraphics[scale=0.7]{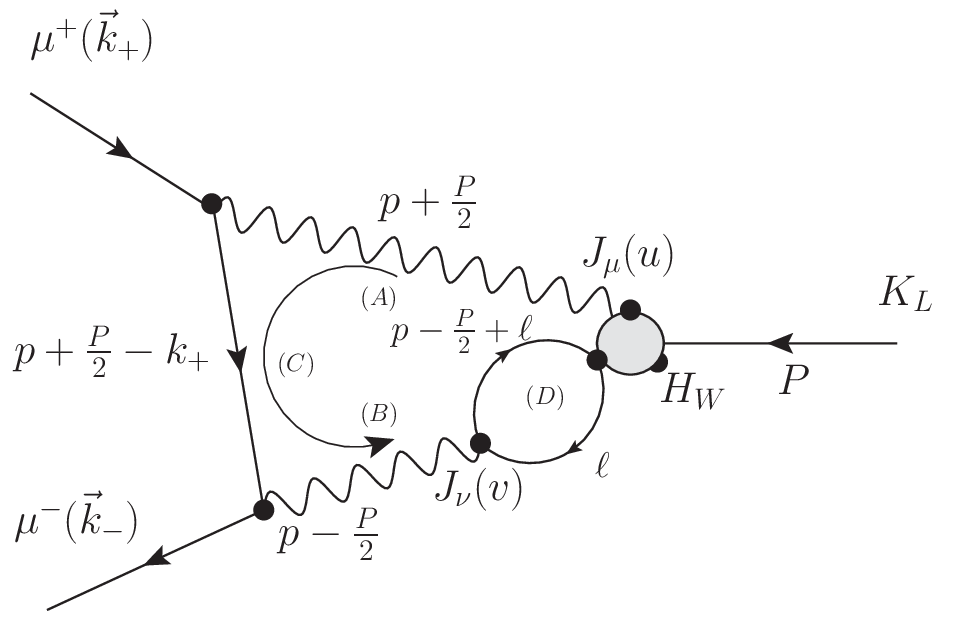}
\caption{The Feynman diagram with a $\pi\pi\gamma$ intermediate state contributing to $\kll$.}
\label{fig:feyn-ppg}
\end{figure}

To estimate the low-energy propagating part of the $\pi\pi\gamma$ contribution to the LD2$\gamma$ decay amplitude, information about the $\kl\rightarrow\pi^+\pi^-\gamma$ and the $\gamma\rightarrow\pi^+\pi^-$ processes is needed.\footnote{It has become common when discussing low-energy processes involving on-shell and virtual photons to add a ``*'' to distinguish a virtual photon. Since here we are discussing both virtual pions and photons and often a photon propagator can contain both virtual and on-shell parts, we do not use * superscripts in the remainder of this paper.}  Assuming CP-conservation, we can parameterize the associated matrix elements using
\begin{equation}\label{eq:fpi}
\lla\pi^+(\textbf{p}^+)\pi^-(\textbf{p}^-)|J_\mu(0)|0\rra
= i(p_+-p_-)_\mu F_\pi^V(s)\,,
\end{equation}
\begin{equation}\label{eq:vpipig}
\begin{split}
& 
\lla \pi^+(\textbf{p}^+)\pi^-(\textbf{p}^-)| J_\nu(0) |\kl \rra
=iV_{K_L\pi\pi\gamma}(s,t)
\epsilon_{\alpha\beta\gamma\nu} (p^+)^\alpha (p^-)^\beta P^\gamma\,,\quad
\end{split}
\end{equation}
where $s\equiv -\left(p^++p^-\right)^2$ and $t \equiv -\left(P-p^+\right)^2$ and we have neglected a second, parity-violating amplitude which is suppressed by additional powers of the pion momenta.  The simplest way to estimate the low-energy contribution of the $\pi\pi\gamma$ state is to use a chiral-perturbation-theory inspired effective model, where the aforementioned processes are described by structure-less point-like interactions.  In such a theory, the two-pion contribution to the LD$2\gamma$ decay amplitude can be represented by the Feynman diagram Fig.~\ref{fig:feyn-ppg}.  In a complete calculation based on this model, the pion-loop integral will be divergent and require regularization and the introduction of a new low energy constant.  Since we are interested in only the low-energy part of the contribution, we regulate the loop integral by putting an upper bound on the energy $E_{\pi\pi}$ of the $\pi\pi$ system in the rest frame of the initial kaon.  The details of such a calculation can be found in Appendix~\ref{sec:amp2pig}.

As described above, we are interested in only a restrictive low-energy region and will assume a momentum-independent, point-like interaction for the $\kl\rightarrow\pi^+\pi^-\gamma$ vertex,
\begin{equation}\label{eq:fpt}
V_{K_L\pi\pi\gamma}(s,t) = V_{K_L\pi\pi\gamma}^\mathrm{pt}\,,
\end{equation}
for our estimate, although a simple effective Lagrangian with vector-meson degrees of freedom~\cite{Lin:1987de} does provide a plausible description of the on-shell $\kl\rightarrow\pi^+\pi^-\gamma$ decay from experiment~\cite{KTeV:2006diq}.  To determine the value of $V_{K_L\pi\pi\gamma}^{\mathrm{pt}}$, we match the decay rate given by this simple point-like model to its experimentally measured value. The decay rate computed from the Dalitz plot density with a generic form factor is given by
\begin{equation}
\Gamma_{\kl\rightarrow\pi\pi\gamma} = \frac{M_K}{64\pi^3}\int_{m_\pi}^{M_K/2}\textbf{p}_+^2 dE_\pi(\textbf{p}_+) \int_{m_\pi}^{M_K/2} \textbf{p}_-^2 dE_\pi(\textbf{p}_-) \left(1-\cos^2\theta\right) |V_{K_L\pi\pi\gamma}(s,t)|^2 \Theta\left( 1-\cos^2\theta\right)\,,
\end{equation}
where 
\begin{equation}
\cos\theta = \frac{M_K^2+2M_\pi^2-2M_K\left(E_\pi(\textbf{p}_+) + E_\pi(\textbf{p}_-)\right)+2E_\pi(\textbf{p}_+)E_\pi(\textbf{p}_-)}{2|\textbf{p}_+||\textbf{p}_-|}\,.
\end{equation}

It should be noted that the $\kl\rightarrow\pi^+\pi^-\gamma$ decay receives both CP-conserving and CP-violating contributions.  The CP-conserving contribution which we need is described as direct emission (DE) while the CP-violating inner bremsstrahlung (IB) contribution does not contribute to the process being studied here~\cite{KTeV:2006diq}.  Experimentally, the ratio between the DE and the total DE+IB has been determined to be 0.689 if only events with photon energy above 20 MeV are included. Using the known branching ratio for this process~\cite{Cirigliano:2011ny}, we find
\begin{equation}
|V_{K_L\pi\pi\gamma}^\mathrm{pt}| = 1.414(0.022)\times 10^{-6} \textrm{ GeV}^{-3}\,.
\end{equation}

\begin{figure}
\includegraphics[scale=0.35]{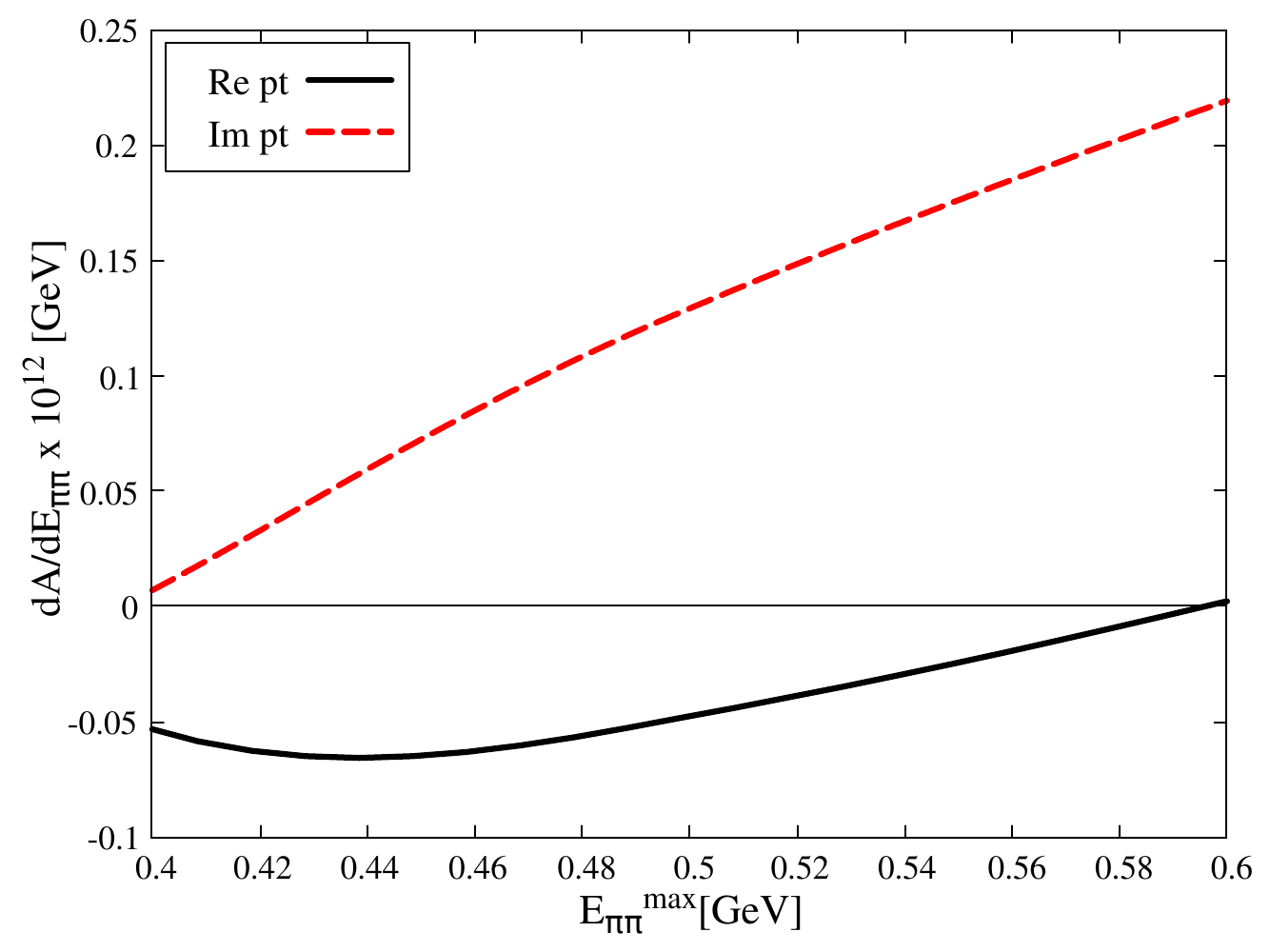}
\includegraphics[scale=0.35]{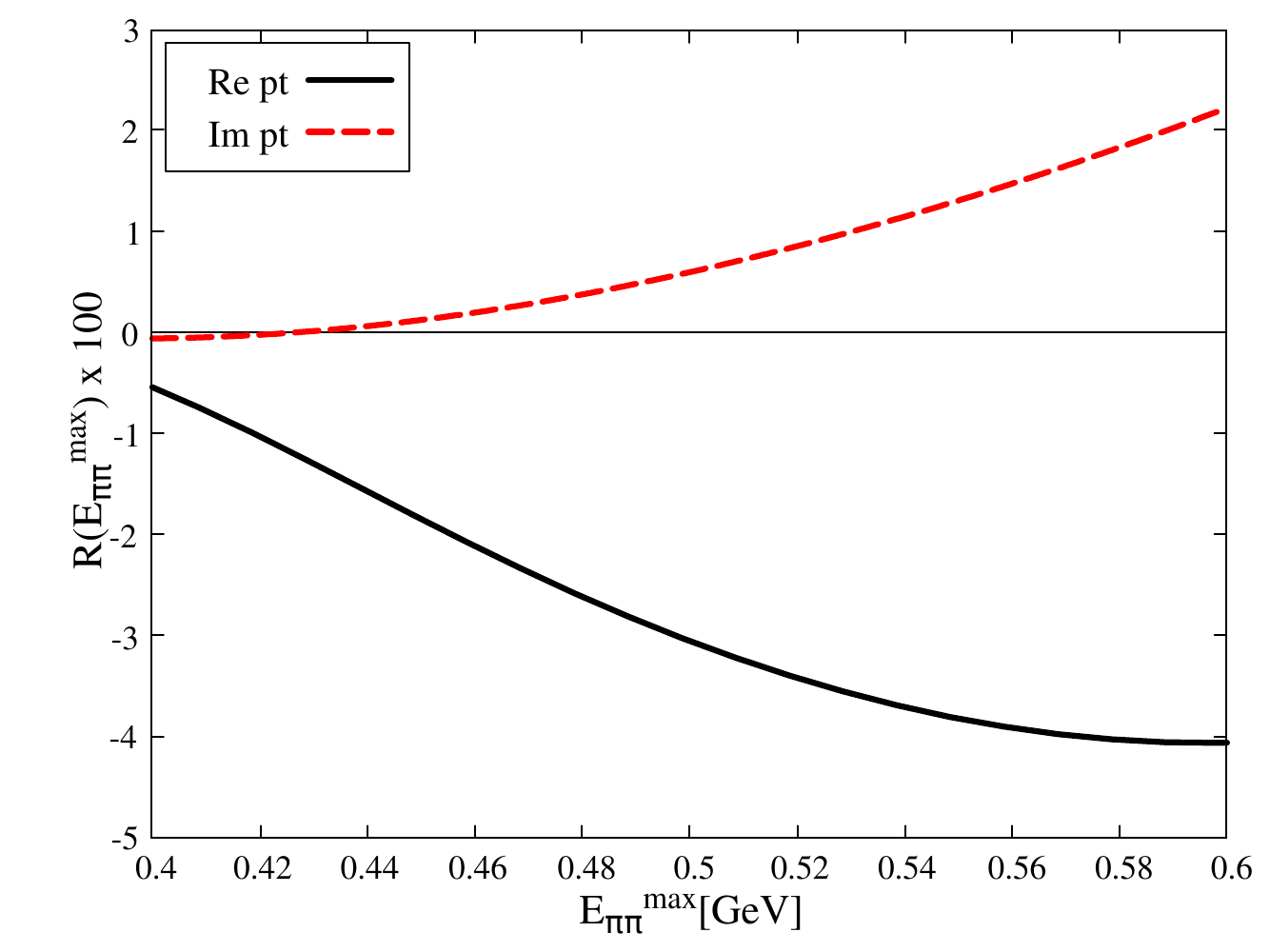}
\caption{Left: The real and imaginary parts of the $\pi\pi\gamma$ amplitude, differential with respect to the $E_{\pi\pi}$, computed using the point-interaction model.  Right: The ratios of the contributions from the low-energy propagating $\pi\pi\gamma$ states to the real and imaginary parts of the KL2$\mu$ decay amplitude estimated in the point-interaction model to their values reconstructed from experiment, plotted as a function of the highest allowed energy for the two-pions in the $K_L$ rest system.  We see effects at the level of 4\% or smaller.}
\label{fig:pipig-ratio-sqed}
\end{figure}

The estimate of the $\pi\pi\gamma$ intermediate state contribution to the LD2$\gamma$ amplitude as a function of the energy cutoff $E_{\pi\pi}^{\rm max}$ with a point-like (scalar-QED) $\pi^+\pi^-\gamma$ coupling, $F_\pi^V(s)=1$, is shown in Fig.~\ref{fig:pipig-ratio-sqed}. In the right panel we plot the ratios of the real and the  imaginary parts of the contributions from the $\pi\pi\gamma$ states with $E_{\pi\pi} \le E_{\pi\pi}^{\rm max}$ to the KL2$\mu$ decay amplitude divided by the corresponding real and imaginary parts of complete decay amplitude determined from the optical theorem and experiment~\cite{Workman:2022ynf}.  We note that the imaginary part coming from on-shell $\pi^+\pi^-\gamma$ states has a sign opposite to that coming from the on-shell $\gamma\gamma$ states.  
As shown in the right panel of the plot, at $E_{\pi\pi}=0.6$ GeV, the fractional contributions from the $\pi\pi$-intermediate state to the real and imaginary parts are given by 
\begin{eqnarray}\label{eq:res-pt}
R_{\mathrm{Re}}^{\mathrm{pt}}(0.6\;\textrm{GeV}) &=& \frac{\mathrm{Re}\left\{ \mathcal{A}_{K_L\mu\mu}^{\pi\pi\gamma\text{-pt}}\right\}}{\mathrm{Re}\left\{ \mathcal{A}_{K_L\mu\mu}\right\} } = -0.041\,, \\
R_{\mathrm{Im}}^{\mathrm{pt}}(0.6\;\textrm{GeV}) &=& \frac{\mathrm{Im}\left\{ \mathcal{A}_{K_L\mu\mu}^{\pi\pi\gamma\text{-pt}}\right\}}{\mathrm{Im}\left\{ \mathcal{A}_{K_L\mu\mu}\right\} } = 0.022\,. \nonumber
\end{eqnarray}

In order to provide information about the $\pi\pi$ energies which give the largest $\pi\pi\gamma$ contributions to the KL2$\mu$ decay amplitude, we plot in the left panel of Fig.~\ref{fig:pipig-ratio-sqed} the $\pi\pi\gamma$ contribution to the real and imaginary parts of the LD$2\gamma$ decay amplitude now differential in $E_{\pi\pi}$.  The imaginary part increases rapidly with increasing $E_{\pi\pi}$ as might be expected from the increasing phase space of the on-shell $\pi\pi\gamma$ intermediate state required by the optical theorem.  However, the real part is the sum of a number of distinct terms with varying signs and a more complicated, non-monotonic behavior.

A difficulty in using this simple model to estimate the $\pi\pi\gamma$ contribution to the LD2$\gamma$ decay amplitude is the sensitivity to the chosen cutoff energy, since the curves grow rapidly with increasing $E_{\pi\pi}$.  We consider the choice of $E_{\pi\pi}=0.6$ GeV as conservative since we are interested in the order of magnitude of the systematic error that results from an inaccurate treatment of the region with $E_{\pi\pi}\lesssim M_K$ in a finite-volume lattice calculation.

To ensure that we do not significantly underestimate this source of systematic error, we will consider a second model, where the effect of the $\rho$ meson is introduced in the form factor $F_\pi^V$.  In nature, the $\rho$ resonance plays an important role in the $\gamma\rightarrow\pi^+\pi^-$ process. A point-like description of this coupling usually leads to much smaller values for observables where long-distance physics is relevant. An example of such is the hadronic vacuum polarization contribution to the anomalous magnetic moment of the muon~\cite{Jegerlehner:2009ry}.  For our purpose, we are concerned with a $\pi\pi$ intermediate state with an energy near that of the kaon.  The inclusion of the $\rho$ resonance would give a reasonable upper bound by enhancing $F_\pi^V$ as a result of the nearby $\rho$ peak with a finite width.  

Of course, the effects of the $\rho$ meson (with a Compton wave length of 0.25 fm) should be accurately reproduced by a lattice calculation performed on a lattice volume volume with a linear extent $\gtrsim 4$ fm.  Thus, we would expect that where such a rho enhancement is significant it does not introduce unusually large finite-volume errors.  However, it may still be of interest to see the effect on our finite-volume error estimate that results from including the enhancement implied by the $\rho$ meson.  

\begin{figure}
\includegraphics[scale=0.35]{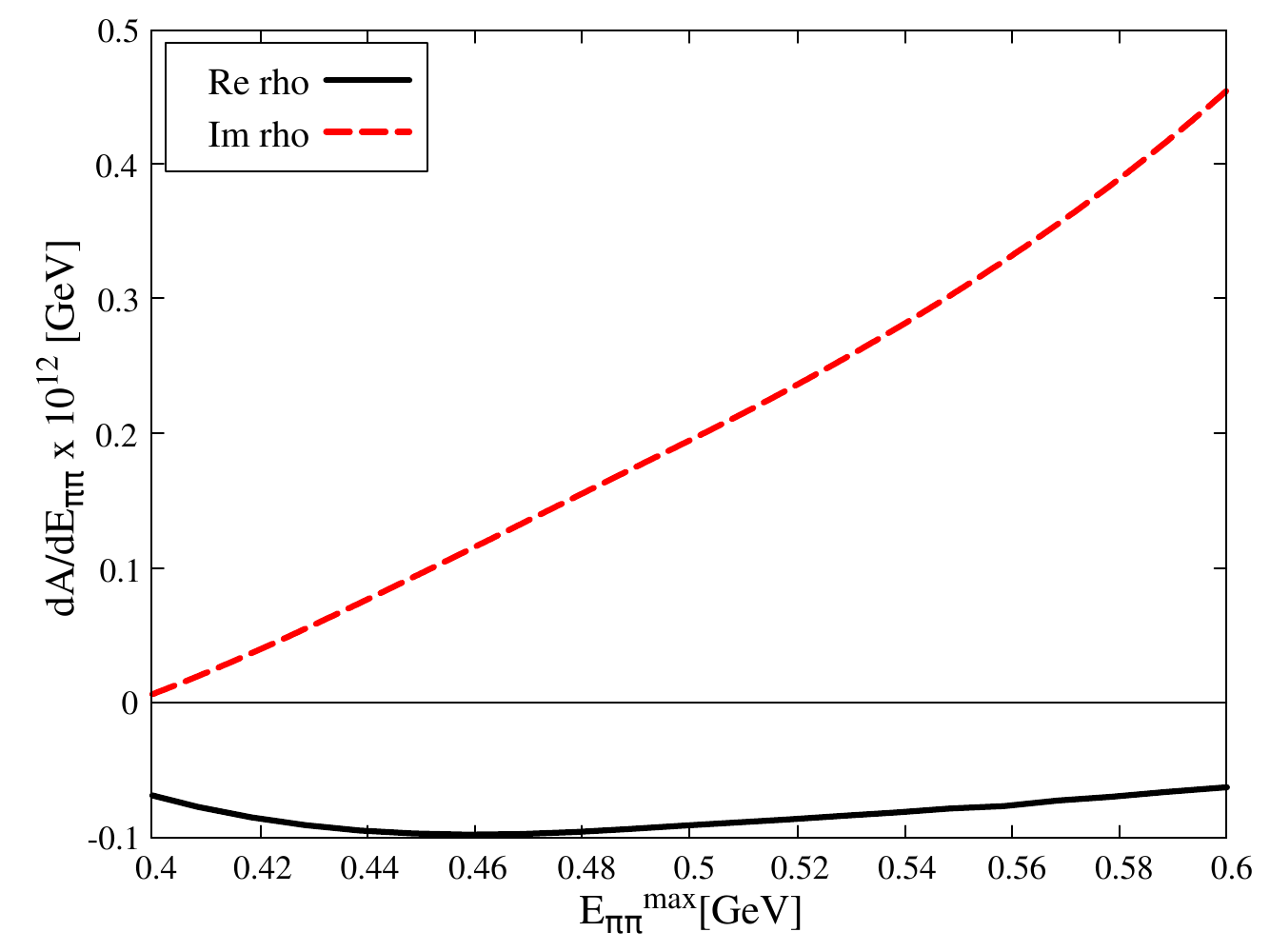}
\includegraphics[scale=0.35]{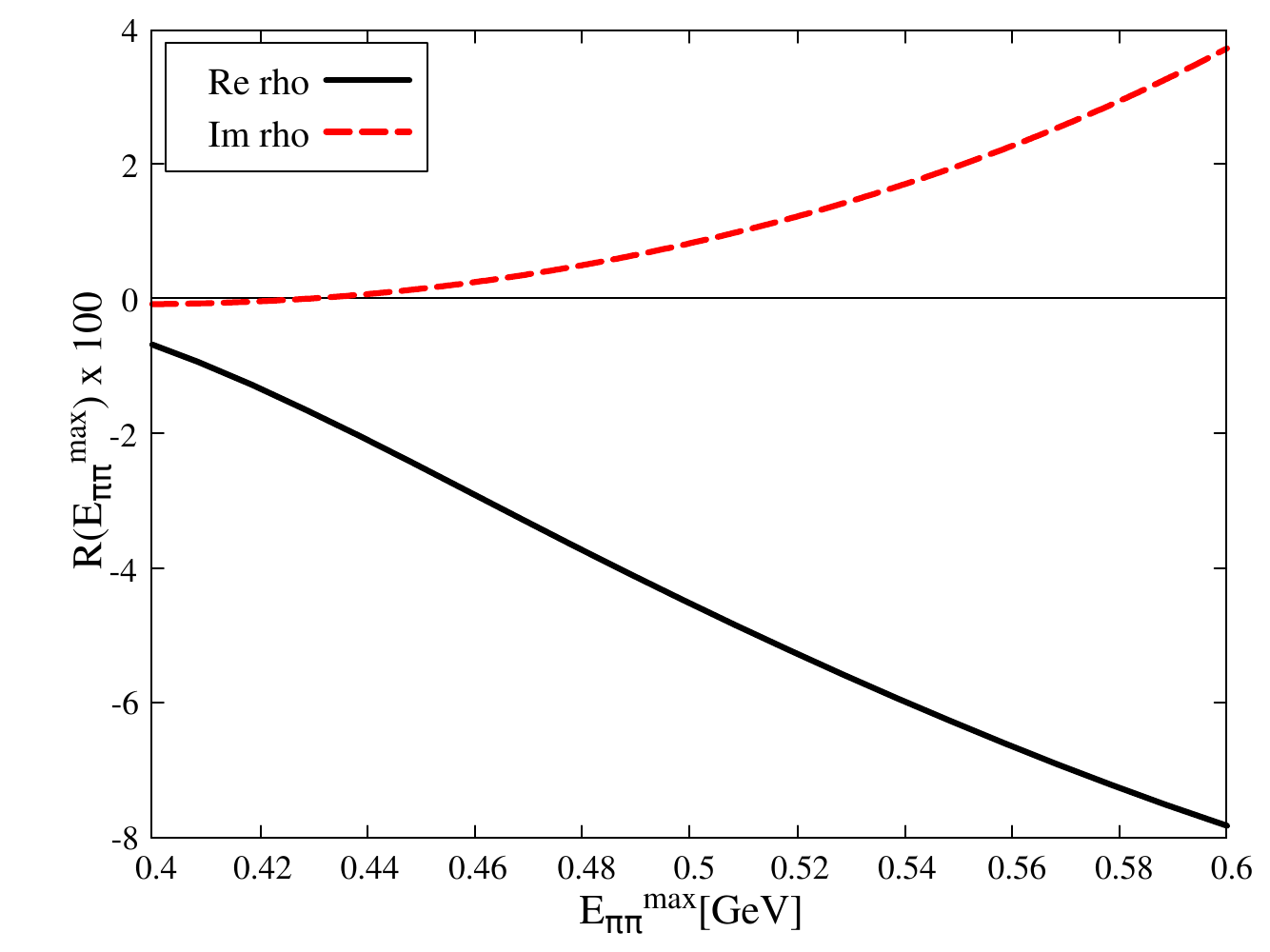}
\caption{The same quantities plotted in Fig.~\ref{fig:pipig-ratio-sqed} but here computed with a Gounaris-Sakuri parametrization of the pion form factor and a point-like $\kl\pi\pi\gamma$ vertex.}
\label{fig:pipig-ratio-gsff}
\end{figure}

As a first approximation, we use the Gounaris-Sakuri (GS) parametrized pion form factor~\cite{Gounaris:1968mw, Francis:2013fzp, Feng:2014gba} for $F_\pi^V$, which takes into account the mass and the width of the $\rho$ meson.  This parametrization gives good predictions for quantities related to the $I=1$ channel $\pi\pi$-scattering.  We obtain the ratios of the real and imaginary parts of the $\pi\pi\gamma$ contribution divided by the experimental values of the total KL2$\mu$ amplitudes as a function of $E_{\pi\pi}^{\mathrm{max}}$ in Fig.~\ref{fig:pipig-ratio-gsff}. At $E_{\pi\pi}=0.6$ GeV, we find

\begin{eqnarray}
R_{\mathrm{Re}}^\rho(0.6\;\textrm{GeV}) &=& \frac{\mathrm{Re}\left\{ \mathcal{A}_{K_L\mu\mu}^{\pi\pi\gamma\text{-}\rho}\right\}}{\mathrm{Re}\left\{ \mathcal{A}_{K_L\mu\mu}\right\} } = -0.078\,, \\
R_{\mathrm{Im}}^\rho(0.6\;\textrm{GeV}) &=& \frac{\mathrm{Im}\left\{ \mathcal{A}_{K_L\mu\mu}^{\pi\pi\gamma\text{-}\rho}\right\}}{\mathrm{Im}\left\{ \mathcal{A}_{K_L\mu\mu}\right\} } = 0.037\,, \nonumber
\end{eqnarray}
for the ratios for the real and the imaginary parts.  Compared to the numbers quoted for the previous case in Eq.~\eqref{eq:res-pt} where $F_\pi^V$ is set to 1, it is reassuring to see that the enhancement is not very significant.

\begin{figure}
\includegraphics[scale=0.5]{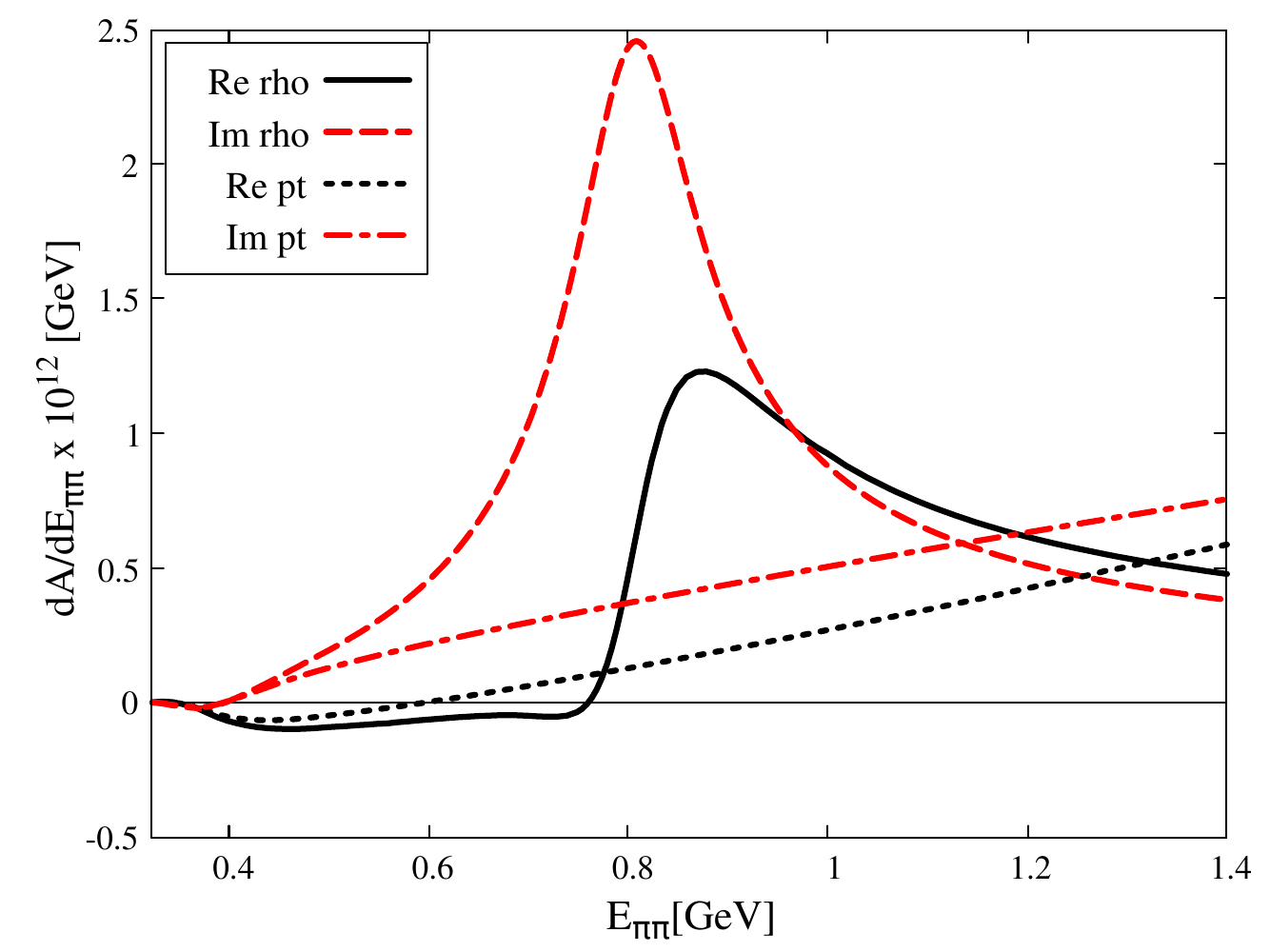}
\caption{Comparison of the $\pi\pi\gamma$ contributions to $K\to\mu\mu$ amplitude differential with respect to the two-pion energy in the lab frame computed with a point-like $K_L\pi\pi\gamma$ vertex and with either a point-like or $\rho$-dominated pion form factor.}
\label{fig:ppg-compare}
\end{figure}

In Fig.~\ref{fig:ppg-compare}, we display the results on the differential amplitude $\frac{d\mathcal{A}_{K_L\mu\mu}^{\pi\pi\gamma\text{-pt}/\rho}(E_{\pi\pi})}{dE_{\pi\pi}}$ obtained from both models over a wider range of $\pi\pi$ energies.  With the inclusion of the $\rho$-resonance, the imaginary part exhibit peaks near the $\rho$-pole. 
It is remarkable that, in the model with the $\rho$, the curve for the real part stays very flat for an extended energy window before suddenly dropping to zero at the resonance. 
Also, one can notice that the slope of the imaginary part in the GS model becomes steeper only when one is within the decay width of the $\rho$.
Consequently, within the low energy window $E_{\pi\pi}^{\rm max}\in[0.4,0.6]$ GeV shown in Figs.~\ref{fig:pipig-ratio-sqed} and \ref{fig:pipig-ratio-gsff}, the imaginary-part from the GS model is comparable to that of the point-like model.
Note that, although the inclusion of the $\rho$ in $F_\pi^V$ does decrease at large $E_{\pi\pi}$, the real part and the imaginary part (coming predominately from the on-shell $\gamma\gamma$) appear not to show the proper asymptotic behavior at $E_{\pi\pi}\gtrsim 1$ GeV.
Indeed, this is expected from the point-like nature of the form factor $V_{K_L\pi\pi\gamma}$.

\section{Conclusion}\label{sec:conclu}

In this paper, we present a framework within which the long-distance two-photon exchange contribution to the complex CP-conserving amplitudes for $\kll$ and $K_{\rm L}\to e^+e^-$ decay can be calculated using a combination of Minkowski-space Feynman amplitudes and Euclidean-space lattice QCD.  Although unphysical contributions appear in the Euclidean-space portion of the calculation, those of importance come from either an intermediate $\pi^0$ or $\eta$ state at rest and can be accurately removed.  The treatment of these states is already familiar in other applications of lattice QCD to flavor physics, {\it e.g.} the calculation of the $\kl$-$K_{\rm S}$ mass difference~\cite{Bai:2014cva}.

Special to these $K_L\to\ell^+\ell^-$ decays is the presence of three-particle $\pi\pi\gamma$ intermediate states with energy smaller than the kaon mass.  Similar to the single pion intermediate state, these low energy $\pi\pi\gamma$ states will  lead to unphysical exponentially growing behavior as the Euclidean time separation of the source and sink in the lattice calculation is increased -- behavior which threatens to obscure the physical part of interest. In contrast with the unphysical contribution of two-pion intermediate states in the $\kl$-$K_{\rm S}$ mass difference calculation~\cite{Christ:2015pwa}, once these unphysical $\pi\pi\gamma$ contributions have been removed it is not known how to restore the physical contribution of these states with controlled finite-volume errors.  Also in contrast with the $\pi\pi$ contribution in the kaon calculation, the favored QED$_\infty$ treatment of electromagnetism in this $\kll$ calculation does permit momentum non-conservation.  This allows the $\kl$ at rest to transition to a combination of a discrete finite-volume $\pi\pi$ state with a photon carrying a continuous three-momentum so that a continuous spectrum of ``finite-volume'' $\pi\pi\gamma$ states must be removed.  

While these would be interesting challenges to overcome, for the purposes of a calculation of $\kll$ decay with 10\% accuracy, we have demonstrated in two model calculations that these troublesome $\pi\pi\gamma$ effects can be safely neglected.  The small size of these effects might be expected given the low-energy suppression coming from both three-body phase space and the appearance of multiple powers for the momenta carried by the two pions and photon in the needed transition amplitudes.  In the model calculation with a point-like $K_L\pi\pi\gamma$ vertex and $\pi\pi\gamma$ vertex presented here the contribution from those $\pi\pi\gamma$ states with $\pi\pi$ energy at or below 600 MeV was found to be 4\% for the dispersive part.  

To give a more complete picture of this process, we presented a second model calculation which included enhancements arising when the effects of the $\rho$ were added to the $\pi\pi\gamma$ vertex.  These were significant increasing the contribution of  $\pi\pi\gamma$ states with $\pi\pi$ energy at or below 600 MeV to 8\%.  However, since we expect that the effects of the $\rho$ will be accurately captured on typical-size lattice volumes, we do not view this 8\% result as a reasonable estimate of the errors likely to result from our failure to accurately reproduce the effects of the $\pi\pi\gamma$ intermediate states.  We conclude that there is a physically important window where the proposed lattice calculation can determine this two-photon-exchange background process to 10\% accuracy, allowing the earlier experimental result for the $\kll$ decay to be used as an accurate test of the short-distance second-order standard model prediction.

An exploratory lattice QCD calculation of this two-photon-exchange process with physical quark masses is now underway~\cite{Chao:2023cxp, Chao:2024cnu} at a single lattice spacing $a$ with $1/a \approx 1$ GeV.  We plan that this first calculation will be followed by calculations at multiple lattice spacings so that a continuum limit can be taken and at multiple physical volumes so that finite-volume errors can be controlled. 

\section*{Acknowledgements}

We thank Ceran Hu for his contributions to the study of the momentum non-conservation resulting from the use of QED$_\infty$ and our RBC and UKQCD Collaboration colleagues for discussion and ideas.  This work was supported in part by the U.S. Department of Energy (DOE) grant \#DE-SC0011941.

\newpage

\appendix

\section{Equality of Minkowski and Euclidean decay amplitudes}\label{sec:M=E}
In this appendix we provide an explicit demonstration of the equality of decay amplitudes computed using Euclidean or Minkowski time dependence provided all of the intermediate states which appear have energies larger than that of the decaying state.  We begin with a general on-shell, order $k+1$ Minkowski-space QCD decay amplitude for a decay from the initial state $|D\rangle$ to the final state $|F\rangle$ with equal energies, $E_D=E_F$ which, exploiting time-translation symmetry, can be write written in the non-covariant form:
\begin{equation}
\mathcal{A}_{F,D,k+1}(t) = \Bigl\langle F\Bigl|\int_0^t dt_k \int_0^{t_k} dt_{k-1}\ldots \int_0^{t_2} dt_1. 
                                            V_{a_k}(t_k)V_{a_{k-1}}(t_{k-1}) \ldots V_{a_1}(t_1) V_{a_0}(0)\Bigr|D\Bigr\rangle.
\label{eq:decay1}
\end{equation}
Here the interaction operators responsible for the decay are written in the Heisenberg representation with their time dependence determined by the QCD Hamiltonian $H_{\mathrm{QCD}}$.  The physical, Minkowski-space decay amplitude can be obtained from $\mathcal{A}_{F,D,k+1}(t)$ in the limit $t\to\infty$.

The amplitude $\mathcal{A}_{F,D,k+1}(t)$ can be built up recursively from lower-order amplitudes $\mathcal{A}_{n_{\ell+1},D,\ell+1}(t)$, also defined by Eq.~\eqref{eq:decay1} if we replace the label $F$ with that for the state $n_{\ell+1}$.  This can be done using the recursion relation:
\begin{eqnarray}
\mathcal{A}_{n_{\ell+1},D,\ell+1}(t_{\ell+1})
                                        &=&\sum_{n_\ell} \int_0^{t_{\ell+1}} dt_\ell \langle n_{\ell+1}|V_{a_\ell}(t_\ell)|n_\ell \rangle 
                                                            \mathcal{A}_{n_\ell,D,\ell}(t_\ell) \nonumber \\
                                       &=& \sum_{n_\ell} \int_0^{t_{\ell+1}} dt_\ell \langle n_{\ell+1}|V_{a_\ell}(0)|n_\ell \rangle 
                                                            \mathcal{A}_{n_\ell,D,\ell}(t_\ell) e^{-i(E_{n_\ell} - E_{n_{\ell+1}}) t_\ell}
\label{eq:recursion}
\end{eqnarray}
and the result for $\mathcal{A}_{2,D,2}(t)$:
\begin{eqnarray}
\mathcal{A}_{n_2,D,2}(t_2)
                                        &=& \sum_{n_1} \int_0^{t_2} dt_1 \langle n_2|V_{a_1}(t_1)|n_1 \rangle 
                                                            \langle n_1|V_{a_0}(0)|D \rangle \nonumber \\
                                       &=&  \sum_{n_1} \langle n_2|V_{a_1}(0)|n_1 \rangle 
                                                            \langle n_1|V_{a_0}(0)|D \rangle 
                                                                       \frac{e^{-i(E_{n_1}-E_{n_2})t_2}-1}{-i(E_{n_1}-E_{n_2})}
\label{eq:start}
\end{eqnarray}

Examining Eqs.~\eqref{eq:recursion} and \eqref{eq:start}, we make the inductive hypothesis that the amplitude $\mathcal{A}_{n_{\ell},D,\ell}(t_\ell)$ depends on the time $t_\ell$ through a sum exponentials of the form $\exp\{-i(E_{n_i} - E_{n_\ell})t_\ell\}$ where $E_{n_i}$ is the energy of one of the members of the complete set of intermediate states that can be inserted between $V_{a_i}(t_i)V_{a_{i-1}}(t_{i-1})$ in the product of operators appearing on the right-hand side of Eq.~\eqref{eq:decay1}.   This hypothesis is easily seen to be obeyed for the lowest-order case $\ell = 2$ from Eq.~\eqref{eq:start}.  The constant "$-1$" in that equation can also be accommodated if we include in the list of energies $E_{n_i}$ that may appear also the energy of the left-most state, $E_{n_\ell}$, so that the behavior $e^{-i(E_{n_i}-E_{n_2})t_2}$ will also include $e^{-i(E_{n_2}-E_{n_2})t_2} = 1$.  The hypothesis is then established by induction if we recognize that the explicit exponential factor which appears in the recursion relation given in Eq.~\eqref{eq:recursion} has the effect of replacing an exponent $-i(E_{n_i}-E_{n_\ell})t_\ell$ that appears in $\mathcal{A}_{n_{\ell},D,\ell}(t_\ell)$ with $(E_{n_i}-E_{n_{\ell+1}})t_{\ell+1}$, demonstrating the inductive hypothesis.

With this understanding of the dependence of $\mathcal{A}_{n_{\ell},D,\ell}(t_\ell)$ on $t_\ell$, we can then consider the integral over $t_k$ defining the decay amplitude in Eq.~\eqref{eq:decay1}, simplified using Eq.~\eqref{eq:recursion}:
\begin{equation}
\mathcal{A}_{F,D,k+1}(t) = \sum_{n_k} \int_0^t dt_k \langle F|V_{a_k}(0)|n_k \rangle 
                                                            \mathcal{A}_{n_k,D,k}(t_k) e^{-i(E_{n_k} - E_F) t_k}.
\label{eq:decay2}
\end{equation}
We have established that as a sum of exponentials the integrand in Eq.~\eqref{eq:decay2} is an analytic function of $t_k$ in the entire complex plane (assuming proper convergence of these sums of exponentials).  We have shown that all of these exponentials have the form $\exp\{-i(E_{n_\ell} - E_F) t_k\}$ where $E_{n_\ell}$ is the energy of a possible intermediate state.  Since by assumption such an energy must be greater than $E_F$, each term in the sum of exponentials that make up the right hand side of Eq.~\eqref{eq:decay2} is exponentially decreasing for large $t_k$ in the fourth quadrant of the complex $t_k$ plane, allowing us to use Cauchy's theorem to Wick rotate the $t_k$ contour from the positive real to the negative imaginary axis.  This establishes the equality of the decay amplitude computed in Minkowski and Euclidean space.   (The lack of explicit exponential suppression for $t_k$ large and real can be remedied in the usual way by replacing the real contour between the points 0 to $|t|$ by the contour connecting 0 and $|t|(1-i\varepsilon)$ or by introducing wave packets and integrating over $E_F$.)

\section{Formula for the low-energy part of the $\pi\pi\gamma$ contribution}\label{sec:amp2pig}

In this appendix, we give more details of the calculation which leads to the estimates for the low-energy part of the $\pi\pi\gamma$ contribution to the LD2$\gamma$ amplitude presented in Sect.~\ref{sec:pi-pi-gamma}.
We will consider the single $s$-wave amplitude obtained by averaging over the directions of the final, out-going muons.
We denote the four-momentum of the initial kaon with $P=\left(M_K,\textbf{0}\right)$ and the four-momenta of the final-state muons with $k^{\pm}=\left(M_K/2, \pm\textbf{k}\right)$, where $\textbf{k}^2=M_K^2/4-m_\mu^2$.
We will adopt the most-plus convention for the metric as in the main text which is more convenient to make a connection to the Euclidean case.

We will assume the parametrization introduced in Eq.~\eqref{eq:fpi} for the pion form factor and the point-like coupling for the $\kl\rightarrow\pi^+\pi^-\gamma$ vertex given in Eq.~\eqref{eq:vpipig} and described in Sect.~\ref{sec:pi-pi-gamma}.  In that case the contribution to the total LD2$\gamma$ $s$-wave decay amplitude from those $\pi\pi\gamma$ intermediate states with a $\pi\pi$ energy in the rest frame of the kaon below $E_{\pi\pi}$ admits a spectral representation over $s$, the invariant-mass-squared of the $\pi\pi$-system: 
\begin{equation}\label{eq:apipi-master}
\mathcal{A}_{K_L\mu\mu}^{\pi\pi\gamma}(E_{\pi\pi}) = 4\pi CM_K^2 \int d^4p\ 
\frac{\textbf{p}^2}{D(p)}\Pi(E_{\pi\pi},p;P)\,,
\end{equation}
where
\begin{equation}\label{eq:pi-master}
\Pi(E_{\pi\pi},p;P)\equiv
\int^{E_{\pi\pi}^2-\textbf{p}^2}_{4M_\pi^2}\frac{ds}{2\pi}\;s\;\eta(s)
[F_\pi^V(s)]^*V^{\rm pt}_{K_L\pi\pi\gamma}
\left[
\frac{2}{\left(p-\frac{1}{2}P\right)^2 +s -i\varepsilon}
\right]
\,,
\end{equation}
\begin{equation}
D(p)\equiv \left[\left(p-\frac{1}{2}P\right)^2-i\varepsilon\right]\left[\left(p+\frac{1}{2}P\right)^2-i\varepsilon\right]\left[\left(p+\frac{1}{2}P-k^+\right)^2+m_\mu^2-i\varepsilon\right]\,,
\end{equation}
\begin{equation}
\eta(s)\equiv \frac{1}{48\pi^2}\left(1-\frac{4M_\pi^2}{s}\right)^{3/2}\,,
\end{equation}
\begin{equation}
C = i\frac{16\pi^2\alpha^2 m_\mu}{(2\pi)^4\sqrt{4\pi\alpha}M_K}\,.
\end{equation}
Some elements of a derivation of Eq.~\eqref{eq:apipi-master} can be found in Refs.~\cite{Schneider:2012ez, Hoferichter:2021lct}.
Note that the phase of $[F_\pi^V(s)]^*$ cancels out that of $V_{\kl\pi\pi\gamma}^{\rm pt}$ as a result of Watson's theorem.

In Sect.~\ref{sec:pi-pi-gamma}, two models for the $\gamma\rightarrow\pi^+\pi^-$ were studied:
in the first case, a point-like scalar-QED-type model was studies where $F_{\pi}^V$ was simply set to unity; 
in the second case, we attempted to include the $\rho$-resonance following the Gounaris-Sakurai parametrization of the pion form factor as in Ref.~\cite{Francis:2013fzp}, with the $\rho$ mass $M_\rho$ and width $\Gamma_\rho$ as inputs.  Specifically we used
\begin{equation}
F_{\pi}^V(s) = \frac{f_0}{\frac{k^3}{\sqrt{s}}\left(\cot\delta_{11}(k)-i\right)}\,,
\end{equation}
where
\begin{equation}
k\equiv \sqrt{\frac{1}{4}s - M_\pi^2}\,,
\end{equation}
and the phase shift is given by the relation
\begin{equation}
\frac{k^3}{\sqrt{s}}\cot\delta_{11}(k) = k^2h(\sqrt{s}) - k_\rho^2h(M_\rho)+b(k^2-k_\rho^2)\,,
\end{equation}
with
\begin{equation}
k_\rho \equiv \sqrt{\frac{1}{4}M_\rho^2-M_\pi^2}\,,
\end{equation}
\begin{equation}
b = - \frac{2}{M_\rho}\left[
\frac{2k_\rho^3}{M_\rho \Gamma_\rho} + \frac{1}{2}M_\rho h(M_\rho) + k_\rho^2h^\prime(M_\rho)
\right]\,,
\end{equation}
\begin{equation}
h(\omega) = \frac{2}{\pi}\frac{k}{\omega}\log\frac{\omega+2k}{2M_\pi}\,,
\end{equation}
\begin{equation}
f_0 = -\frac{M_\pi^2}{\pi}-k_\rho^2 h(M_\rho)-b\frac{M_\rho^2}{4}\,.
\end{equation}
For our numerical applications, we used $M_\rho = 775$ MeV and $\Gamma_\rho= 130$ MeV.

\bibliography{references}

\end{document}